\documentclass[superscriptaddress,prl,floatfix,twocolumn,amsmath,amssymb,aps,longbibliography, footinbib]{revtex4-2}
\usepackage{graphicx}
\usepackage{amsmath}
\usepackage{amssymb}
\usepackage{color}
\usepackage{multirow}
\usepackage{bm}
\usepackage{subfigure}
\usepackage{tabularx}
\usepackage{hhline}  
\newcolumntype{d}[1]{D{.}{.}{#1}}

\begin{document}
	\title{Switching Chern number by sliding and gating in alternately twisted tetralayer MoTe$_2$}
	\author{Xiao-Wei Zhang}
	\affiliation{Department of Materials Science and Engineering, University of Washington, Seattle, WA 98195, USA}
	\author{Kaijie Yang}
	\affiliation{Department of Materials Science and Engineering, University of Washington, Seattle, WA 98195, USA} 
    \author{Xiaodong Xu}
    \affiliation{Department of Materials Science and Engineering, University of Washington, Seattle, WA 98195, USA}
	\affiliation{Department of Physics, University of Washington, Seattle, WA 98195, USA}
	\author{Ting Cao}
	\email{tingcao@uw.edu}
	\affiliation{Department of Materials Science and Engineering, University of Washington, Seattle, WA 98195, USA}
	\author{Di Xiao}
	\email{dixiao@uw.edu}
	\affiliation{Department of Materials Science and Engineering, University of Washington, Seattle, WA 98195, USA}
	\affiliation{Department of Physics, University of Washington, Seattle, WA 98195, USA}
	\date{\today}
\begin{abstract}
 Switching the bulk Chern number in topological materials is of central importance for the design of topological electronic devices. 
Motivated by recent observations of integer and fractional quantum anomalous Hall effects in twisted transition metal dichalcogenides (tTMDs), we realize the switching of valley Chern number through sliding and gating in alternately twisted tetralayer (ATT) MoTe$_{2}$. 
Using large-scale density functional theory (DFT) calculations, we show that the Chern number of the first $K$-valley moir\'e band evolves from $+1$ to $-1$ under the interlayer sliding. 
Furthermore, an applied electric field can switch the valley Chern number from $-1$ to $+1$. 
Based on the developed continuum model, we reveal that these switching behaviors are caused by the sliding- and gate-dependent intralayer moir\'e potential distributions across the layers.
Our results establish ATT MoTe$_{2}$ as a promising platform for engineering moir\'e band topologies through the design of moir\'e potentials with sliding in multilayer moir\'e systems. 

\end{abstract}

\maketitle

\textit{Introduction}---In a Chern insulator, the bulk-boundary correspondence dictates that the sign of the bulk Chern number determines the chirality
or propagating direction of edge currents. The ability to switch the bulk Chern number is therefore of central importance for the design of next-generation topological electronic devices\cite{gilbert2021topological,chang2023colloquium,yuan2024electrical}.
In recent years, moiré TMD superlattices, such as tWSe$_2$ and tMoTe$_2$ bilayers, have emerged as such appealing platforms in which integer and fractional quantum anomalous Hall effects have been observed in the absence of an external magnetic field\cite{cai2023signatures,zeng2023thermodynamic,park2023observation,xu2023observation,foutty2024mapping}.

Currently, there are mainly two routes to realize the switching of bulk Chern numbers for the first moir\'e band in these systems: controlling the valley polarization (the occupation imbalance of doped carriers between the K/K$^{'}$ valleys)~\cite{cai2023signatures,park2023observation,xu2023observation,zeng2023thermodynamic,cai2026optical,huber2026optical,holtzmann2026optical}, or
controlling the valley Chern number~\cite{cai2023signatures,park2023observation,xu2023observation,zeng2023thermodynamic,foutty2024mapping, zhang2024polarization,morales2023pressure,jiao2026hydrostatic,ding2026sliding,zheng2024interlayer,liang2025moire,nakatsuji2025moire,choi2025higher,fan2026layerwise,qi2026chern}. 
The first route relies on the fact that the frontier moir\'e mini bands from the two valleys carry opposite valley Chern numbers~\cite{li2026quantum}. 
The valley polarization can be reversed by applying a magnetic field, thereby switching the Chern number~\cite{cai2023signatures,zeng2023thermodynamic,park2023observation,xu2023observation}.
More recently, optical control of valley polarization has been demonstrated via hole-exciton interactions in tMoTe$_{2}$ bilayers at filling factors of $-1$ and $-2/3$, enabling Chern number switching~\cite{cai2026optical,huber2026optical,holtzmann2026optical}.
For the second route, recent large-scale DFT calculations predicted that, in tWSe$_{2}$ bilayers, the valley Chern numbers of the first moiré band can be switched by tuning the twist angle 
from $\sim$4$^\circ$ to 1$^\circ$~\cite{zhang2024polarization}. However, tuning the twist angle over such a wide range is challenging in experiments.
Beyond the twist angle, other tuning knobs have also been used to control the valley Chern number, e.g., electric field~\cite{cai2023signatures,park2023observation,xu2023observation,zeng2023thermodynamic,foutty2024mapping}, pressure~\cite{jiao2026hydrostatic,morales2023pressure}, interlayer sliding~\cite{zheng2024interlayer,ding2026sliding}, and  stacking order~\cite{jia2024moire,liang2025moire,nakatsuji2025moire,choi2025higher,fan2026layerwise,qi2026chern}. Nevertheless, these approaches have so far realized only the transition between C=$\pm1$ and C=$0$; in other words, the turning on and off of band topology.
Therefore, switching the valley Chern number for the first moir\'e band at a fixed twist angle remains a challenge.

\begin{figure}[t]
	\includegraphics[width=1.0\columnwidth]{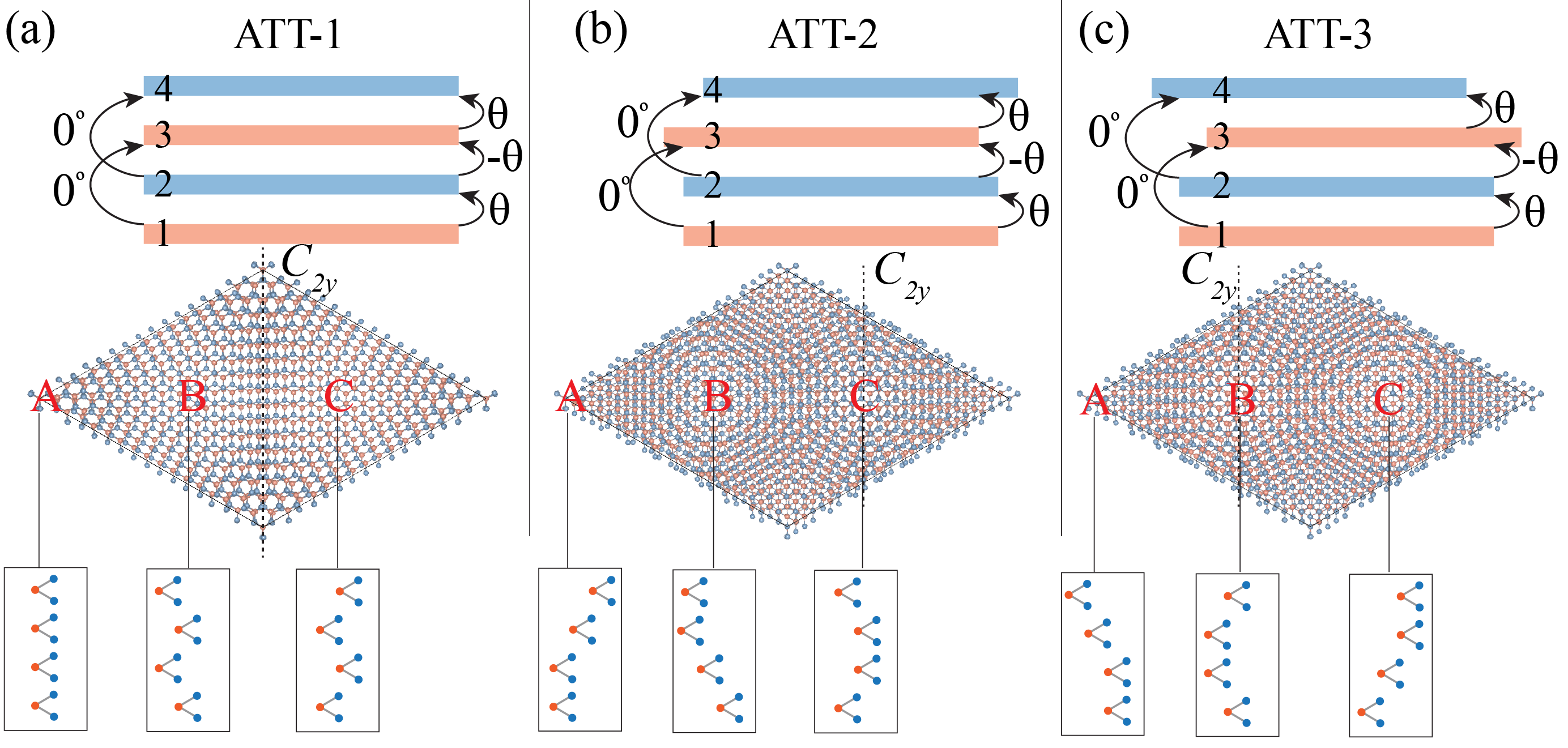}
		\caption{Schematic illustration of the three sliding configurations of ATT MoTe$_{2}$, (a)-(c).
		The four layers are alternately rotated with a twist angle sequence $(\theta, -\theta, \theta)$. Layer 1 (2)
		is aligned with layer 3 (4), as denoted by 0$^\circ$ between them. 
		In (a), ATT-1, layers 1 (2) and 3 (4) exhibit MM stacking. In (b), ATT-2, layers 1 (2) and 3 (4) exhibit MX (XM) stacking. In (c), ATT-3, 
		layers 1 (2) and 3 (4) exhibit XM (MX) stacking. Here, ``M'' denotes the Mo atom while ``X'' denotes the Te atom, and MX denotes that M is positioned on top of X.
		 The vertical dashed line denotes the $C_{2y}$ axis, which shifts upon sliding. 
		The bottom boxes show the local stacking structures at the three high symmetry sites A, B, and C. 
			 }
		\label{fig1}
	\end{figure}
In this Letter, we demonstrate the switching of valley Chern numbers in ATT MoTe$_{2}$. 
Using large-scale DFT calculations, we show that the Chern number of the first $K$-valley moiré band evolves from $+1$ to 0, 
and then to $-1$ across three considered sliding configurations [Fig.~\ref{fig1}]. 
This switching behavior is understood within a continuum model analysis, which reveals that the intralayer moir\'e potential is 
significantly reshaped upon sliding. Moreover, we show that, for one of the sliding configurations, an applied electric field  
can switch the valley Chern number of the first moiré band from $-1$ to $+1$. 
Our results establish ATT MoTe$_{2}$ as a promising platform for engineering moiré band topologies and demonstrate sliding as an effective tuning knob in moir\'e multilayer systems.

\textit{Sliding structure}---Figure 1 schematically illustrates three sliding configurations of ATT MoTe$_{2}$. Other possible sliding configurations are not considered here, as they do not exhibit Chern-number switching.
In the following, we denote the three structures as ATT-1, ATT-2, and ATT-3. The four layers are alternately rotated with a $(\theta, -\theta, \theta)$ sequence, giving rise to the same moir\'e periodicity as that of a $\theta$ twisted MoTe$_{2}$ bilayer. 
In contrast to the bilayer case, however, ATT structures possess an additional 
interlayer sliding degree of freedom, resulting in richer structural tunability. 
Specifically, layer 1(2) is aligned or untwisted with respect to layer 3(4), and sliding layers 3 and 4 with respect to layers 1 and 2 generates inequivalent configurations. 
In ATT-1, layers 1(2) and 3(4) exhibit MM (MM) stacking, whereas in ATT-2, layers 1 (2) and 3(4) exhibit MX (XM) stacking. 
In ATT-3, layers 3 and 4 are slid in the opposite direction relative to ATT-2, resulting in XM (MX) stacking between layers 1(2) and 3(4). 
In addition, sliding produces different combinations of local bilayer stackings around the three $C_{3z}$ (threefold rotation) high symmetry sites, as shown in the bottom boxes of Fig.~\ref{fig1}.
These stacking variations can have important effects on moir\'e potentials, as discussed below. Notably, all three configurations preserve $C_{2y}$ symmetry (twofold rotation about the $y$ axis), but the $C_{2y}$ axis passes through different high 
symmetry sites for different sliding configurations. 

Using machine learning force fields (MLFFs), we efficiently perform moiré-scale lattice relaxations for the three configurations. 
The MLFFs are parameterized using the deep potential molecular dynamics (DeePMD) framework~\cite{wang2018deepmd,zeng2023deepmd}, with training data generated from $ab$ $initio$ molecular dynamics simulations performed using the VASP package~\cite{kresse1996efficiency}. 
Further details on the MLFF parameterization and structural relaxations are provided in Sections I and II of the Supplemental Material~\cite{supp}.
After relaxations, we find that ATT-1 has the lowest energy, while ATT-2 and ATT-3 are slightly higher in energy by $\sim$3 meV/atom. 
This small energy difference is consistent with weak van der Waals interlayer interactions.

\begin{figure}[t]
	\includegraphics[width=1.0\columnwidth]{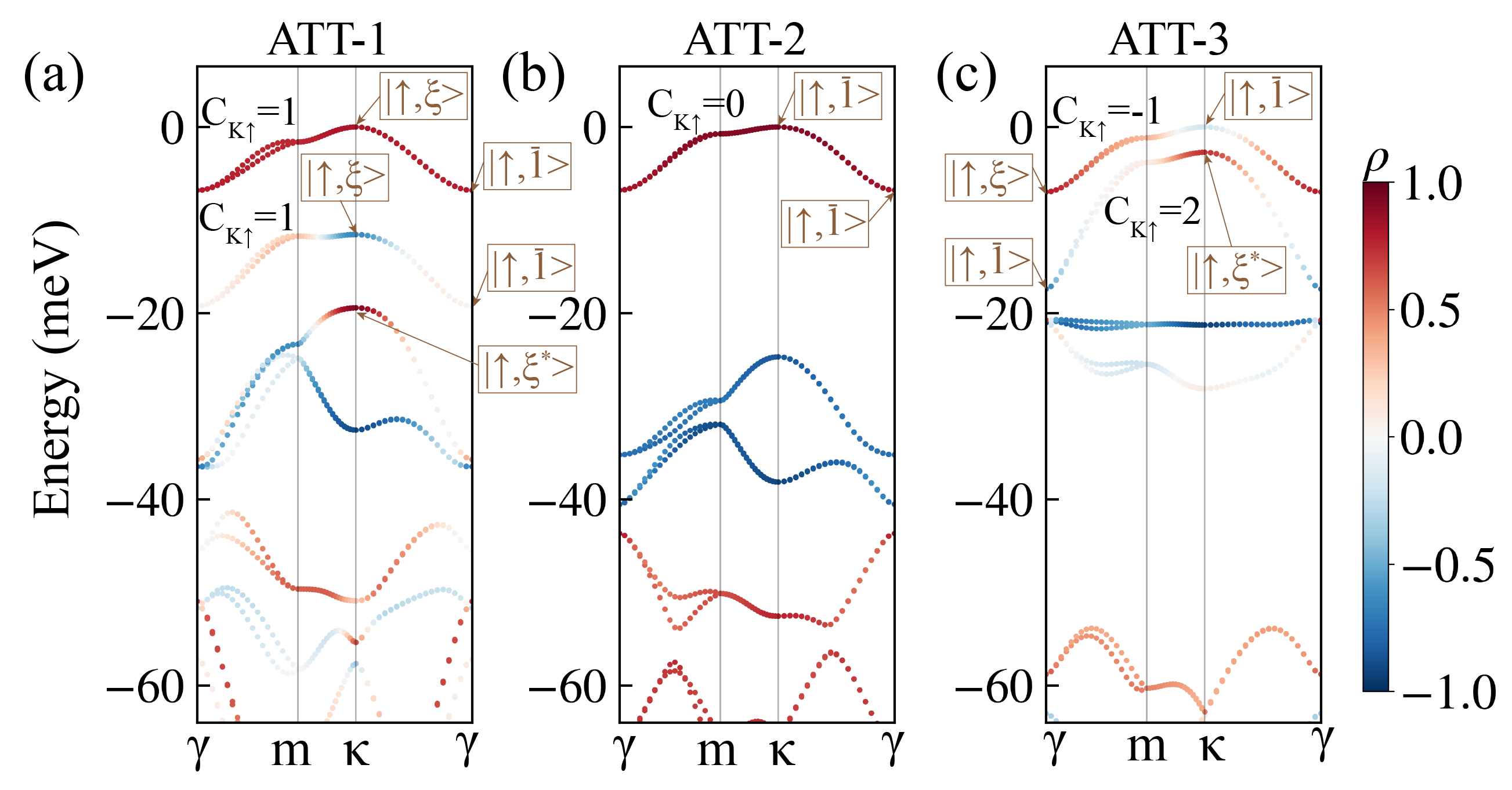}
		\caption{ Large-scale DFT band structures for the three sliding configurations, (a)-(c). 
		The color denotes the layer polarization $\rho$ of the moir\'e wavefunctions. Owing to $C_{2y}$ symmetry, layers 1 (2) and 4 (3) have the same polarization weights. 
		Positive (negative) values of $\rho$ indicate that the wave functions are more polarized in layers 2 and 3 (layers 1 and 4). 
		The Chern numbers of the spin-up $K$-valley moir\'e bands are indicated. The brown boxes show the spin-ful $C_{3z}$ eigenvalues for the spin-up bands at the $\gamma$ and $\kappa$ points, where $\xi$=$e^{i\pi/3}$, $\xi^{\ast}$=$e^{-i\pi/3}$, and $\bar{1}$=$-1$.
		The $C_{3z}$ eigenvalue at the $\kappa'$ point is the same as at the $\kappa$ point. 
		}
		\label{fig2}
	\end{figure}

\textit{Chern number switching by sliding}---With the relaxed structures, we perform large-scale DFT calculations to obtain the moiré band structures. 
The calculations are carried out using the SIESTA package~\cite{soler2002siesta}, and further details are provided in Section III of the Supplemental Material~\cite{supp}. 
Figure 2 shows the valence moiré band structures for the three sliding configurations with a twist-angle sequence $(3.89^\circ,-3.89^\circ,3.89^\circ)$.
We find that sliding significantly modifies the band structures. Importantly, the Chern number of the first spin-up $K$-valley moiré band evolves from $+1$ to $0$, 
and then to $-1$ upon sliding. In addition, for ATT-1, the second band carries Chern number $+1$, opposite to that of the tMoTe$_2$ bilayer at the same twist angle~\cite{wang2024fractional,zhang2024polarization}. 
We further note that, for ATT-3, the second $K$-valley moiré band carries Chern number $+2$, which is not found in twisted bilayers~\cite{zhang2024polarization,jia2024moire}.

\begin{figure*}[t]
	\includegraphics[width=1.0\linewidth]{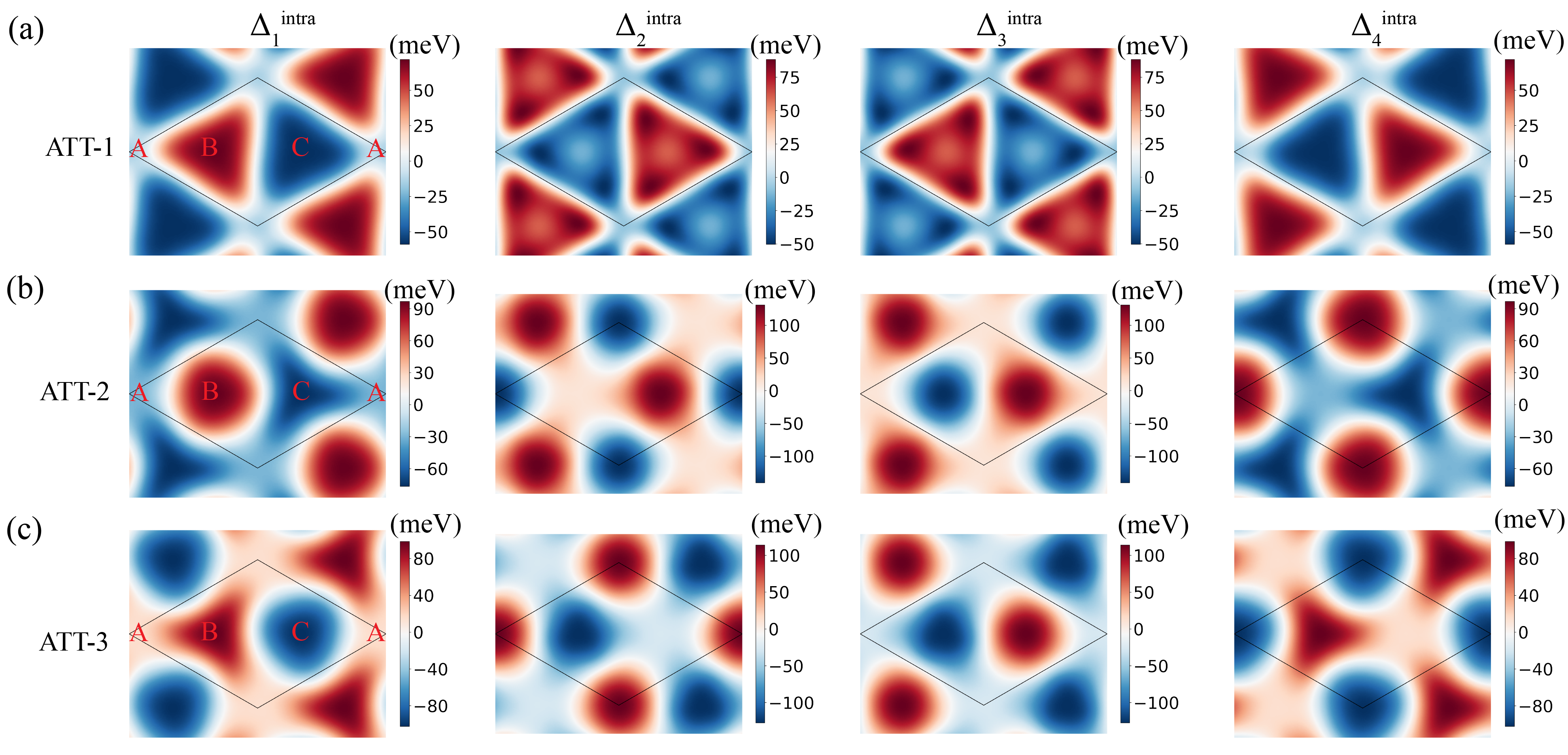}
	\caption{ Distributions of the intralayer moir\'e potential $\Delta_{i}^{\text{intra}}$ for the three sliding configurations, (a)-(c). From left to right, the four columns correspond to the four layers. 
	The potential distribution in layer 1 (2) is related to that in layer 4 (3) by $C_{2y}$ symmetry. The three high-symmetry sites are labeled. 
	The black parallelogram denotes a moir\'e unit cell. 
	}
	\label{fig3}
\end{figure*}

To understand the band topology evolution induced by sliding, we further analyze the layer polarization of moir\'e wave functions in Fig.~\ref{fig2}. 
Since the system preserves $C_{2y}$ symmetry, the layer polarization $\rho$ is defined as 
\begin{equation}
	\rho = \frac{\int{d\bm{r}}\sum_{i=2,3}|\phi_{i,n\bm{k}}|^2 - \int{d\bm{r}}\sum_{i=1,4}|\phi_{i,n\bm{k}}|^2 }{\int{d\bm{r}}\sum_{i=2,3}|\phi_{i,n\bm{k}}|^2 + \int{d\bm{r}}\sum_{i=1,4}|\phi_{i,n\bm{k}}|^2 },
\end{equation}
where $i$ and $n$ denote the layer and band index, respectively. 
For ATT-1, we find that the first band is predominantly polarized in layers 2 and 3, 
while the second band and third bands exhibit band inversion around the $\kappa/\kappa'$ points, as evidenced by their layer polarizations. 
For comparison, we construct a reference structure by separating layers 2 and 3 from the relaxed ATT structure and compute its band structures, shown in Fig.S3 of Ref.\cite{supp}. 
We denote this structure as ATT-1mm with ``mm'' representing ``middle moir\'e''. 
We find that the first two bands of ATT-1mm carry Chern numbers $(+1, -1)$, which is the same as that of tMoTe$_{2}$ bilayer at 3.89$^\circ$~\cite{zhang2024polarization,wang2024fractional}. 
By tracking the evolution of $C_{3z}$ eigenvalues, we identify that the band inversion between the inner and outer layers around the $\kappa/\kappa'$ points causes the C=$-1$ to C=$+1$ transition. 

For ATT-2, the first band is also mainly polarized in layers 2 and 3, while the second band mainly originates from layers 1 and 4. 
Compared with the reference ATT-2mm structure (see Fig.S3 of Ref.~\cite{supp}), the Chern number of the first band remains unchanged. 
However, because the second band of ATT-2mm is well separated from the first band, in the case of ATT-2, the second and even third bands are predominantly contributed by layers 1 and 4, while the fourth band is mainly polarized in layers 2 and 3.

For ATT-3, the first two bands are predominantly polarized in layers 2 and 3, except near the $\gamma$ point for the second band. 
Compared with ATT-3mm (see Fig. S3 of Ref.~\cite{supp}), the Chern numbers of both the first and second bands are modified. 
This change originates from band inversion between the first and second bands at the $\kappa/\kappa'$ points, leading to a Chern number $-1$ for the first band. 
In addition, in the tetralayer case, further band inversions between the second, third, and fourth bands at the $\gamma$ point give rise to a Chern number $+2$ for the second band.
These comparisons clearly indicate that interlayer hybridization between the inner and outer layers plays a central role in driving the evolution of band topology upon sliding. 
To understand how sliding modifies interlayer hybridizations and induces band inversions, we next turn to a continuum model analysis.

\textit{Continuum model}---We first generalize the recently developed twist-angle transferable continuum model for twisted bilayers~\cite{zhang2025twist} to the tetralayer case. 
The total Hamiltonian for the spin-up $K$-valley is written as 
\begin{widetext}
\begin{equation}
  \mathcal{H}_{K \uparrow} = \left(\begin{array}{c}
      - \frac{(\bm{p} - \hbar \bm{\kappa}_1 + e\bm{A}_1)^2} {2 m^{\ast}} + h_{\text{kin,1}}^{h.o.}+  \Delta_{1}^{\text{intra}}(\bm{r})  \hspace*{\fill}
    \Tilde{\Delta}_{T_{12}} (\bm{r}, \bm{k})   \hspace{4em}\hspace*{\fill}  \hspace{4em}\hspace*{\fill}   \\
    \Tilde{\Delta}_{T_{12}}^{\dagger} (\bm{r}, \bm{k}) \hspace{2em}\hspace*{\fill}   - \frac{(\bm{p} - \hbar \bm{\kappa}_2 + e\bm{A}_2)^2}{2m^{\ast}} + h_{\text{kin,2}}^{h.o.}+ \Delta_2^{\text{intra}}(\bm{r})  \hspace{2em}\hspace*{\fill}  \Tilde{\Delta}_{T_{23}} (\bm{r}, \bm{k})  \hspace{6em}\hspace*{\fill} \\ 
    \hspace{8em}\hspace*{\fill}   \Tilde{\Delta}_{T_{23}}^{\dagger} (\bm{r}, \bm{k})   \hspace{2em}\hspace*{\fill}   - \frac{(\bm{p} - \hbar \bm{\kappa}_3 + e\bm{A}_3)^2} {2 m^{\ast}} + h_{\text{kin,3}}^{h.o.}+ \Delta_3^{\text{intra}}(\bm{r})  \hspace{2em}\hspace*{\fill}  \Tilde{\Delta}_{T_{34}} (\bm{r}, \bm{k}) \hspace{2em} \hspace*{\fill} \\ 
	\hspace{8em}\hspace*{\fill}  \hspace*{\fill} \Tilde{\Delta}_{T_{34}}^\dagger (\bm{r}, \bm{k})   \hspace*{\fill}       - \frac{(\bm{p} - \hbar \bm{\kappa}_4 + e\bm{A}_4)^2} {2 m^{\ast}} + h_{\text{kin,4}}^{h.o.} + \Delta_4^{\text{intra}}(\bm{r})
  \end{array}\right),
  \label{total}
\end{equation}
\end{widetext}
where $\bm{\kappa}_{i}$ denotes the momentum shift induced by twisting, $\bm{A}_{i}$ the strain-induced vector potential, and
$h_{kin, i}^{h.o.}$ the higher-order kinetic energy terms.  
$\Delta^\text{intra}_{i}$ denotes the intralayer potential and includes the following terms, 
\begin{equation}
    \Delta^{\text{intra}_{i}}(\bm{r}) = V_{\varepsilon,i}(\bm{r}) + V_{f,i}(\bm{r}) + V_{p,i}(\bm{r}) + V_{s,i}(\bm{r}) + V_{0}\delta_{i=2,3}.
\end{equation}
Here, $V_{\varepsilon,i}$ originally denotes an average potential between the layers arising from local stacking variations in the twisted bilayer case, $V_{f,i}$ the charge-transfer-induced ferroelectric potential, 
$V_{p,i}$ the strain-gradient-induced piezoelectric potential, and $V_{s,i}$ the strain-induced scalar potential.  The inclusion of the piezoelectric potential is important as it has been demonstrated to have a large effect on the band topology~\cite{zhang2024polarization}. 
$\Tilde{\Delta}_{T_{ii+1}}$ denotes the interlayer tunneling between adjacent layers. 
We note that there exists local $C_{2y}$ symmetry breaking between layers 1 and 2 (similarly between layers 3 and 4). 
To capture this effect, we introduce two additional parameters modifying $V_{\varepsilon,i}$ and $V_{f,i}$, as well as a constant potential $V_{0}$. 
All other parameters are rigidly transferred from the twisted bilayer to the tetralayer system. 
The three additional parameters are determined by fitting to DFT bands of the sliding structures. 
Further details on the derivations of the continuum model and parameter fitting are provided in Section IV of the Supplemental Materials~\cite{supp}.

\textit{Mori\'e potential}---Figure 3 plots the layer-resolved intralayer moiré potential $\Delta^{\mathrm{intra}}_i$ for the three sliding configurations. 
We find that sliding strongly reshapes the moiré potential landscape. Most notably, within a given layer, the locations of potential maxima and minima shift upon sliding, 
except for layer 1 (since layers 1 and 2 are fixed).
In particular, because layer 4 is related to layer 1 by $C_{2y}$ symmetry and the $C_{2y}$ axis itself shifts upon sliding, the potential maximum (minimum) in layer 4 moves from the C (B) site to the A (C) site, and then to the B (A) site across the three configurations. 
Similar shifts of potential extrema also occur in layers 2 and 3.
Compared with ATT-1, in ATT-2 the potential maximum (minimum) in layer 2 shifts to the C (A) site, while in layer 3 it shifts to the C (B) site. 
In ATT-3, the potential maximum in layer 2 shifts to the A (B) site, while in layer 3 it shifts to the C (B) site.

The shifts of potential maxima and minima can be understood by analyzing the local stacking configurations around the high-symmetry sites. 
First, we find that the intralayer moiré potential is dominated by the piezoelectric potential contribution (see Fig. S6 of Ref.~\cite{supp}). 
Second, the sign of the piezoelectric potential can be inferred from the local stacking configurations at such large twist angles~\cite{zhang2024polarization,zhang2025twist}. 
For instance, in twisted bilayers, around MX and XM sites the piezoelectric potential is positive on the M side and negative on the X side, while it nearly vanishes at MM sites.
In the tetralayer case, for example, in ATT-2, because the local stacking at the C site is XMMX, the piezoelectric potential is positive around the C site in layers 2 and 3, and negative around the same site in layers 1 and 4.
We further note that in ATT-1, three potential maxima appear surrounding the C (B) site and three minima appear surrounding the B (C) site in layer 2 (3). 
This behavior originates from stronger in-plane lattice relaxations around the B and C sites in layers 2 and 3 compared with layers 1 and 4. 
As a result, the stacking domains around these sites become more uniform, which reduces the piezoelectric potential in these regions (see additional details in Figs. S2 and S6 of Ref.~\cite{supp}).

The rearrangement of potential maxima is responsible for the evolution of band topology upon sliding. 
For ATT-1, the potential maxima in layers 2 and 3 form an approximate honeycomb lattice, which underlies the non-trivial topology for the first moir\'e band. 
In addition, because the potential maxima in layers 1 and 4 are in energy close to those in layers 2 and 3, the hybridizations between the inner and outer layers induce band inversion between the second and third bands at the $\kappa/\kappa'$ points. 
For ATT-2, the potential maxima in layers 2 and 3 form a triangular lattice, leading to a trivial first moir\'e band.
Moreover, because the potential maximum difference between the inner and outer layers is large, the second band (mainly polarized in layers 1 and 4) is well separated from the first band (mainly polarized in layers 2 and 3). 

Compared with ATT-1 and ATT-2, the situation in ATT-3 is more subtle. The potential maxima in layers 3 and 2 also form a honeycomb lattice, giving rise to the non-zero Chern number of the first moir\'e band. 
However, because now the potential maximum in layer 3 aligns with the C site of layer 2 where the potential nearly vanishes, the interlayer potential difference between layers 2 and 3
becomes smaller compared with ATT-1. In addition, the band gap between the first two bands near the $\kappa/\kappa'$ points is primarily controlled by the interlayer potential difference. 
As a result, the two bands become close to each other around the $\kappa/\kappa'$ points and are easily inverted due to the interlayer hybridizations between the inner and outer layers, leading to a reversal of the Chern number compared with ATT-1.

\begin{figure}[t]
	\includegraphics[width=1.0\columnwidth]{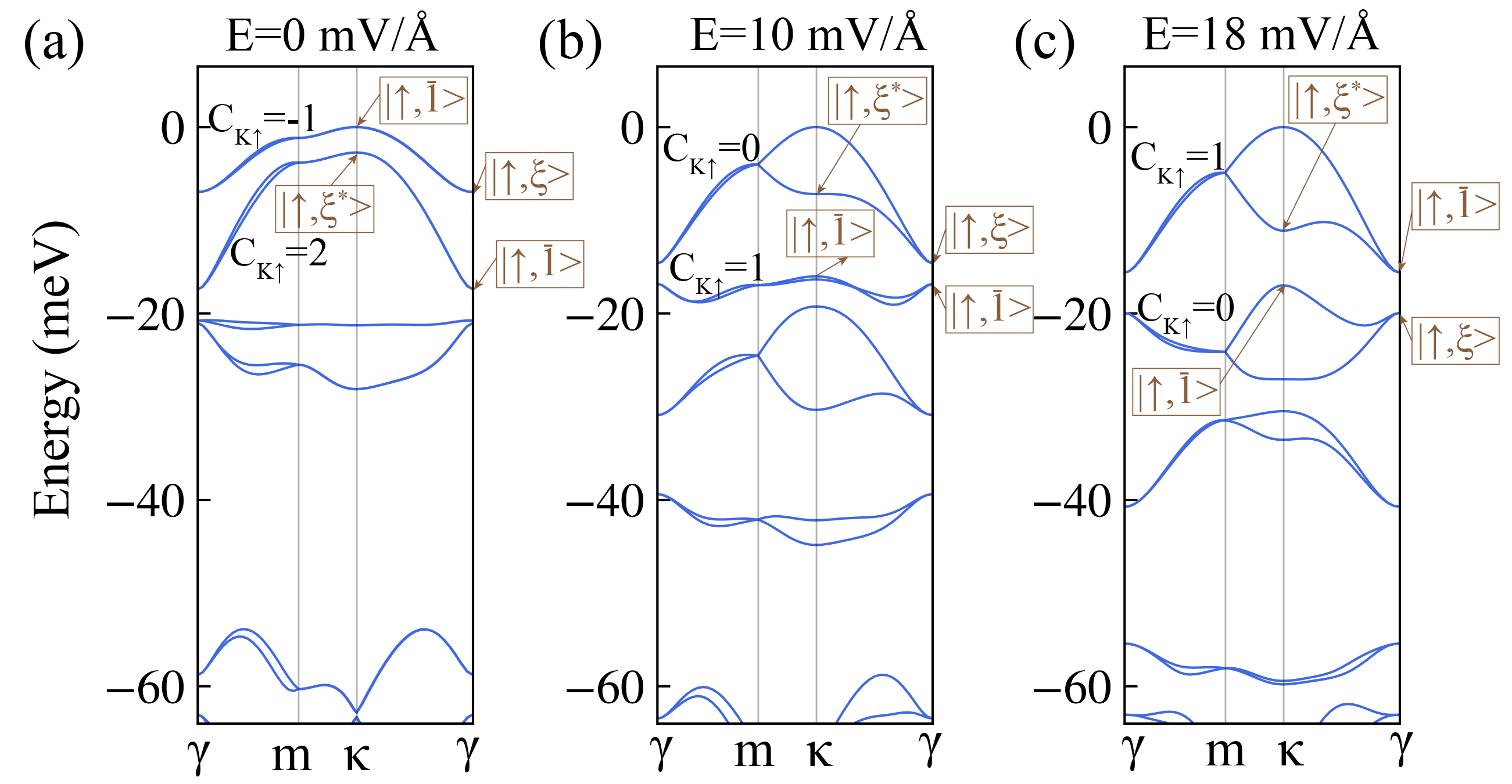}
	\caption{ Large-scale DFT moir\'e band structures of ATT-3 MoTe$_{2}$ at 3.89 $^\circ$ under different electric fields, (a), $E$=0 mV/$\text{\AA}$; (b), $E$=10 mV/$\text{\AA}$; (c), $E$=18 mV/$\text{\AA}$.
	The Chern numbers of the first two spin-up $K$-valley moir\'e bands are labeled. The brown boxes give the spin-ful $C_{3z}$ eigenvalues for the spin-up bands at the $\gamma$ and $\kappa$ points, where $\xi$=$e^{i\pi/3}$, $\xi^{\ast}$=$e^{-i\pi/3}$, and $\bar{1}$=$-1$.
	}
	\label{fig4}
\end{figure}  

\textit{Chern number switching by gating}---We now use an electric field to switch the valley Chern number. 
Figure~\ref{fig4} shows the large-scale DFT band structures of ATT-3 under different electric fields at 3.89$^\circ$. 
It can be seen that as the electric field $E$ increases, the $K$-valley Chern number of the first moir\'e band changes from $-1$ to 0, and then to $-1$. 
Note that due to the presence of $C_{2y}$ symmetry at zero field, a negative electric field produces the same switching behavior of Chern number. 
The change of the Chern number can be understood by tracking the 
$C_{3z}$ eigenvalues. As $E$ increases from 0 to 10 mV/$\text{\AA}$, the first and second spin-up bands are inverted at the $\kappa$ point but not at the $\kappa'$ point, 
and the Chern number of the first band thus becomes 0.  
As $E$ further increases to 18 mV/$\text{\AA}$, an additional inversion occurs at the $\gamma$ point between the first and second spin-up bands, resulting in a Chern number $+1$ for the first band. 

These band inversions and Chern number evolutions are again caused by the electric-field-dependent intralayer moir\'e potential distributions. To capture this effect, we add the following gate potential into Eq.~\ref{total},
	\begin{equation}
	  \mathcal{H}_{E} = \left(\begin{array}{c}
		  -V_{E}/2  \hspace{3em}\hspace*{\fill} \hspace{3em}\hspace*{\fill}  \hspace{3em}\hspace*{\fill}   \\
		 \hspace{3em}\hspace*{\fill}   -V_{E}/4  \hspace{3em}\hspace*{\fill}    \hspace{3em}\hspace*{\fill} \\ 
		\hspace{3em}\hspace*{\fill}     \hspace{3em}\hspace*{\fill}    V_{E}/4  \hspace{3em}\hspace*{\fill}   \\ 
		\hspace{3em}\hspace*{\fill}  \hspace{3em}\hspace*{\fill}   \hspace{3em}\hspace*{\fill}       V_{E}/2
	  \end{array}\right).
	  \label{VE}
	\end{equation}
As shown in Fig.S7 of Ref.~\cite{supp}, at a large gate potential ($V_{E}=42$ meV), the potential maxima shift to layers 3 and 4, and they form a honeycomb lattice, resembling the twisted bilayer at 3.89$^\circ$. 
Consequently, the Chern number of the first moiré band becomes $+1$. 

\textit{Discussion}---In summary, we demonstrate the switching of valley Chern numbers in ATT MoTe$_{2}$ at 3.89$^\circ$. 
Using the interlayer sliding, the Chern number of the first moir\'e band can be switched from $+1$ to $-1$. 
For ATT-3, an experimentally accessible electric field can further switch the Chern number of the first band from $-1$ to $+1$. 
These switching behaviors originate from the reshaped intralayer moir\'e potential distributions across the four layers under sliding and electric fields. 
For experimental realization, interlayer sliding is naturally present in multilayer fabrication processes, rendering the above sliding configurations experimentally feasible. 
Alternatively, various sliding configurations can be obtained through the formation of a supermoiré lattice by tuning the twist-angle sequence.
This finding of switching Chern number establishes ATT MoTe$_{2}$ as a promising platform for engineering moir\'e band topologies and exploring novel quantum states.
For example, in experiments by controlling the sliding or realizing a tetralayer supermoir\'e, 
it is likely to realize the so-called Chern mosaic~\cite{shi2021moire,grover2022chern,devakul2023magic,zhang2024manipulation, xia2025topological,fujimoto2025four} or topological mosaic~\cite{tong2017topological} with mesoscopically distributed topological invariants, 
where the existing gapless states along domain walls are promising for the design of topological electric circuits~\cite{ovchinnikov2022topological,zhao2023creation,yan2024rules,gilbert2025chern,bhattacharjee2026mesoscopic}.

\textit{Acknowledgements}---X.-W.Zhang thanks Huiyuan Zheng, Naiyuan J. Zhang, and Weijie Li for stimulating discussions. The work is mainly supported by the University of Washington Molecular Engineering Materials Center at the University of Washington (DMR-2308979).  The development of machine-learning-enabled methods and advanced codes was supported by the Computational Materials Sciences Program funded by the U.S. Department of Energy, Office of Science, Basic Energy Sciences, Materials Sciences, and Engineering Division, PNNL FWP 83557. 
This research used resources of the National Energy Research Scientific Computing Center, a DOE Office of Science User Facility supported by the Office of Science of the U.S. Department of Energy under Contract No. DE-AC02-05CH11231 using NERSC award BES-ERCAP0037097, BES-ERCAP0037104, and BES-ERCAP0035954.
This work was also facilitated through the use of advanced computational, storage, and networking infrastructure provided by the Hyak supercomputer system. 


\begin{thebibliography}{50}%
	\makeatletter
	\providecommand \@ifxundefined [1]{%
	 \@ifx{#1\undefined}
	}%
	\providecommand \@ifnum [1]{%
	 \ifnum #1\expandafter \@firstoftwo
	 \else \expandafter \@secondoftwo
	 \fi
	}%
	\providecommand \@ifx [1]{%
	 \ifx #1\expandafter \@firstoftwo
	 \else \expandafter \@secondoftwo
	 \fi
	}%
	\providecommand \natexlab [1]{#1}%
	\providecommand \enquote  [1]{``#1''}%
	\providecommand \bibnamefont  [1]{#1}%
	\providecommand \bibfnamefont [1]{#1}%
	\providecommand \citenamefont [1]{#1}%
	\providecommand \href@noop [0]{\@secondoftwo}%
	\providecommand \href [0]{\begingroup \@sanitize@url \@href}%
	\providecommand \@href[1]{\@@startlink{#1}\@@href}%
	\providecommand \@@href[1]{\endgroup#1\@@endlink}%
	\providecommand \@sanitize@url [0]{\catcode `\\12\catcode `\$12\catcode `\&12\catcode `\#12\catcode `\^12\catcode `\_12\catcode `\%12\relax}%
	\providecommand \@@startlink[1]{}%
	\providecommand \@@endlink[0]{}%
	\providecommand \url  [0]{\begingroup\@sanitize@url \@url }%
	\providecommand \@url [1]{\endgroup\@href {#1}{\urlprefix }}%
	\providecommand \urlprefix  [0]{URL }%
	\providecommand \Eprint [0]{\href }%
	\providecommand \doibase [0]{https://doi.org/}%
	\providecommand \selectlanguage [0]{\@gobble}%
	\providecommand \bibinfo  [0]{\@secondoftwo}%
	\providecommand \bibfield  [0]{\@secondoftwo}%
	\providecommand \translation [1]{[#1]}%
	\providecommand \BibitemOpen [0]{}%
	\providecommand \bibitemStop [0]{}%
	\providecommand \bibitemNoStop [0]{.\EOS\space}%
	\providecommand \EOS [0]{\spacefactor3000\relax}%
	\providecommand \BibitemShut  [1]{\csname bibitem#1\endcsname}%
	\let\auto@bib@innerbib\@empty
	\bibitem [{\citenamefont {Gilbert}(2021)}]{gilbert2021topological}%
	  \BibitemOpen
	  \bibfield  {author} {\bibinfo {author} {\bibfnamefont {M.~J.}\ \bibnamefont {Gilbert}},\ }\href@noop {} {\bibfield  {journal} {\bibinfo  {journal} {Communications Physics}\ }\textbf {\bibinfo {volume} {4}},\ \bibinfo {pages} {70} (\bibinfo {year} {2021})}\BibitemShut {NoStop}%
	\bibitem [{\citenamefont {Chang}\ \emph {et~al.}(2023)\citenamefont {Chang}, \citenamefont {Liu},\ and\ \citenamefont {MacDonald}}]{chang2023colloquium}%
	  \BibitemOpen
	  \bibfield  {author} {\bibinfo {author} {\bibfnamefont {C.-Z.}\ \bibnamefont {Chang}}, \bibinfo {author} {\bibfnamefont {C.-X.}\ \bibnamefont {Liu}},\ and\ \bibinfo {author} {\bibfnamefont {A.~H.}\ \bibnamefont {MacDonald}},\ }\href@noop {} {\bibfield  {journal} {\bibinfo  {journal} {Reviews of Modern Physics}\ }\textbf {\bibinfo {volume} {95}},\ \bibinfo {pages} {011002} (\bibinfo {year} {2023})}\BibitemShut {NoStop}%
	\bibitem [{\citenamefont {Yuan}\ \emph {et~al.}(2024)\citenamefont {Yuan}, \citenamefont {Zhou}, \citenamefont {Yang}, \citenamefont {Zhao}, \citenamefont {Zhang}, \citenamefont {Yan}, \citenamefont {Zhuo}, \citenamefont {Mei}, \citenamefont {Wang}, \citenamefont {Yi} \emph {et~al.}}]{yuan2024electrical}%
	  \BibitemOpen
	  \bibfield  {author} {\bibinfo {author} {\bibfnamefont {W.}~\bibnamefont {Yuan}}, \bibinfo {author} {\bibfnamefont {L.-J.}\ \bibnamefont {Zhou}}, \bibinfo {author} {\bibfnamefont {K.}~\bibnamefont {Yang}}, \bibinfo {author} {\bibfnamefont {Y.-F.}\ \bibnamefont {Zhao}}, \bibinfo {author} {\bibfnamefont {R.}~\bibnamefont {Zhang}}, \bibinfo {author} {\bibfnamefont {Z.}~\bibnamefont {Yan}}, \bibinfo {author} {\bibfnamefont {D.}~\bibnamefont {Zhuo}}, \bibinfo {author} {\bibfnamefont {R.}~\bibnamefont {Mei}}, \bibinfo {author} {\bibfnamefont {Y.}~\bibnamefont {Wang}}, \bibinfo {author} {\bibfnamefont {H.}~\bibnamefont {Yi}}, \emph {et~al.},\ }\href@noop {} {\bibfield  {journal} {\bibinfo  {journal} {Nature Materials}\ }\textbf {\bibinfo {volume} {23}},\ \bibinfo {pages} {58} (\bibinfo {year} {2024})}\BibitemShut {NoStop}%
	\bibitem [{\citenamefont {Cai}\ \emph {et~al.}(2023)\citenamefont {Cai}, \citenamefont {Anderson}, \citenamefont {Wang}, \citenamefont {Zhang}, \citenamefont {Liu}, \citenamefont {Holtzmann}, \citenamefont {Zhang}, \citenamefont {Fan}, \citenamefont {Taniguchi}, \citenamefont {Watanabe} \emph {et~al.}}]{cai2023signatures}%
	  \BibitemOpen
	  \bibfield  {author} {\bibinfo {author} {\bibfnamefont {J.}~\bibnamefont {Cai}}, \bibinfo {author} {\bibfnamefont {E.}~\bibnamefont {Anderson}}, \bibinfo {author} {\bibfnamefont {C.}~\bibnamefont {Wang}}, \bibinfo {author} {\bibfnamefont {X.}~\bibnamefont {Zhang}}, \bibinfo {author} {\bibfnamefont {X.}~\bibnamefont {Liu}}, \bibinfo {author} {\bibfnamefont {W.}~\bibnamefont {Holtzmann}}, \bibinfo {author} {\bibfnamefont {Y.}~\bibnamefont {Zhang}}, \bibinfo {author} {\bibfnamefont {F.}~\bibnamefont {Fan}}, \bibinfo {author} {\bibfnamefont {T.}~\bibnamefont {Taniguchi}}, \bibinfo {author} {\bibfnamefont {K.}~\bibnamefont {Watanabe}}, \emph {et~al.},\ }\href@noop {} {\bibfield  {journal} {\bibinfo  {journal} {Nature}\ }\textbf {\bibinfo {volume} {622}},\ \bibinfo {pages} {63} (\bibinfo {year} {2023})}\BibitemShut {NoStop}%
	\bibitem [{\citenamefont {Zeng}\ \emph {et~al.}(2023{\natexlab{a}})\citenamefont {Zeng}, \citenamefont {Xia}, \citenamefont {Kang}, \citenamefont {Zhu}, \citenamefont {Kn{\"u}ppel}, \citenamefont {Vaswani}, \citenamefont {Watanabe}, \citenamefont {Taniguchi}, \citenamefont {Mak},\ and\ \citenamefont {Shan}}]{zeng2023thermodynamic}%
	  \BibitemOpen
	  \bibfield  {author} {\bibinfo {author} {\bibfnamefont {Y.}~\bibnamefont {Zeng}}, \bibinfo {author} {\bibfnamefont {Z.}~\bibnamefont {Xia}}, \bibinfo {author} {\bibfnamefont {K.}~\bibnamefont {Kang}}, \bibinfo {author} {\bibfnamefont {J.}~\bibnamefont {Zhu}}, \bibinfo {author} {\bibfnamefont {P.}~\bibnamefont {Kn{\"u}ppel}}, \bibinfo {author} {\bibfnamefont {C.}~\bibnamefont {Vaswani}}, \bibinfo {author} {\bibfnamefont {K.}~\bibnamefont {Watanabe}}, \bibinfo {author} {\bibfnamefont {T.}~\bibnamefont {Taniguchi}}, \bibinfo {author} {\bibfnamefont {K.~F.}\ \bibnamefont {Mak}},\ and\ \bibinfo {author} {\bibfnamefont {J.}~\bibnamefont {Shan}},\ }\href@noop {} {\bibfield  {journal} {\bibinfo  {journal} {Nature}\ }\textbf {\bibinfo {volume} {622}},\ \bibinfo {pages} {69} (\bibinfo {year} {2023}{\natexlab{a}})}\BibitemShut {NoStop}%
	\bibitem [{\citenamefont {Park}\ \emph {et~al.}(2023)\citenamefont {Park}, \citenamefont {Cai}, \citenamefont {Anderson}, \citenamefont {Zhang}, \citenamefont {Zhu}, \citenamefont {Liu}, \citenamefont {Wang}, \citenamefont {Holtzmann}, \citenamefont {Hu}, \citenamefont {Liu} \emph {et~al.}}]{park2023observation}%
	  \BibitemOpen
	  \bibfield  {author} {\bibinfo {author} {\bibfnamefont {H.}~\bibnamefont {Park}}, \bibinfo {author} {\bibfnamefont {J.}~\bibnamefont {Cai}}, \bibinfo {author} {\bibfnamefont {E.}~\bibnamefont {Anderson}}, \bibinfo {author} {\bibfnamefont {Y.}~\bibnamefont {Zhang}}, \bibinfo {author} {\bibfnamefont {J.}~\bibnamefont {Zhu}}, \bibinfo {author} {\bibfnamefont {X.}~\bibnamefont {Liu}}, \bibinfo {author} {\bibfnamefont {C.}~\bibnamefont {Wang}}, \bibinfo {author} {\bibfnamefont {W.}~\bibnamefont {Holtzmann}}, \bibinfo {author} {\bibfnamefont {C.}~\bibnamefont {Hu}}, \bibinfo {author} {\bibfnamefont {Z.}~\bibnamefont {Liu}}, \emph {et~al.},\ }\href@noop {} {\bibfield  {journal} {\bibinfo  {journal} {Nature}\ }\textbf {\bibinfo {volume} {622}},\ \bibinfo {pages} {74} (\bibinfo {year} {2023})}\BibitemShut {NoStop}%
	\bibitem [{\citenamefont {Xu}\ \emph {et~al.}(2023)\citenamefont {Xu}, \citenamefont {Sun}, \citenamefont {Jia}, \citenamefont {Liu}, \citenamefont {Xu}, \citenamefont {Li}, \citenamefont {Gu}, \citenamefont {Watanabe}, \citenamefont {Taniguchi}, \citenamefont {Tong} \emph {et~al.}}]{xu2023observation}%
	  \BibitemOpen
	  \bibfield  {author} {\bibinfo {author} {\bibfnamefont {F.}~\bibnamefont {Xu}}, \bibinfo {author} {\bibfnamefont {Z.}~\bibnamefont {Sun}}, \bibinfo {author} {\bibfnamefont {T.}~\bibnamefont {Jia}}, \bibinfo {author} {\bibfnamefont {C.}~\bibnamefont {Liu}}, \bibinfo {author} {\bibfnamefont {C.}~\bibnamefont {Xu}}, \bibinfo {author} {\bibfnamefont {C.}~\bibnamefont {Li}}, \bibinfo {author} {\bibfnamefont {Y.}~\bibnamefont {Gu}}, \bibinfo {author} {\bibfnamefont {K.}~\bibnamefont {Watanabe}}, \bibinfo {author} {\bibfnamefont {T.}~\bibnamefont {Taniguchi}}, \bibinfo {author} {\bibfnamefont {B.}~\bibnamefont {Tong}}, \emph {et~al.},\ }\href@noop {} {\bibfield  {journal} {\bibinfo  {journal} {Physical Review X}\ }\textbf {\bibinfo {volume} {13}},\ \bibinfo {pages} {031037} (\bibinfo {year} {2023})}\BibitemShut {NoStop}%
	\bibitem [{\citenamefont {Foutty}\ \emph {et~al.}(2024)\citenamefont {Foutty}, \citenamefont {Kometter}, \citenamefont {Devakul}, \citenamefont {Reddy}, \citenamefont {Watanabe}, \citenamefont {Taniguchi}, \citenamefont {Fu},\ and\ \citenamefont {Feldman}}]{foutty2024mapping}%
	  \BibitemOpen
	  \bibfield  {author} {\bibinfo {author} {\bibfnamefont {B.~A.}\ \bibnamefont {Foutty}}, \bibinfo {author} {\bibfnamefont {C.~R.}\ \bibnamefont {Kometter}}, \bibinfo {author} {\bibfnamefont {T.}~\bibnamefont {Devakul}}, \bibinfo {author} {\bibfnamefont {A.~P.}\ \bibnamefont {Reddy}}, \bibinfo {author} {\bibfnamefont {K.}~\bibnamefont {Watanabe}}, \bibinfo {author} {\bibfnamefont {T.}~\bibnamefont {Taniguchi}}, \bibinfo {author} {\bibfnamefont {L.}~\bibnamefont {Fu}},\ and\ \bibinfo {author} {\bibfnamefont {B.~E.}\ \bibnamefont {Feldman}},\ }\href@noop {} {\bibfield  {journal} {\bibinfo  {journal} {Science}\ }\textbf {\bibinfo {volume} {384}},\ \bibinfo {pages} {343} (\bibinfo {year} {2024})}\BibitemShut {NoStop}%
	\bibitem [{\citenamefont {Cai}\ \emph {et~al.}(2026)\citenamefont {Cai}, \citenamefont {Pan}, \citenamefont {Wang}, \citenamefont {Rasmita}, \citenamefont {Yang}, \citenamefont {Zhao}, \citenamefont {Wang}, \citenamefont {Duan}, \citenamefont {He}, \citenamefont {Watanabe} \emph {et~al.}}]{cai2026optical}%
	  \BibitemOpen
	  \bibfield  {author} {\bibinfo {author} {\bibfnamefont {X.}~\bibnamefont {Cai}}, \bibinfo {author} {\bibfnamefont {H.}~\bibnamefont {Pan}}, \bibinfo {author} {\bibfnamefont {Y.}~\bibnamefont {Wang}}, \bibinfo {author} {\bibfnamefont {A.}~\bibnamefont {Rasmita}}, \bibinfo {author} {\bibfnamefont {S.}~\bibnamefont {Yang}}, \bibinfo {author} {\bibfnamefont {Y.}~\bibnamefont {Zhao}}, \bibinfo {author} {\bibfnamefont {W.}~\bibnamefont {Wang}}, \bibinfo {author} {\bibfnamefont {R.}~\bibnamefont {Duan}}, \bibinfo {author} {\bibfnamefont {R.}~\bibnamefont {He}}, \bibinfo {author} {\bibfnamefont {K.}~\bibnamefont {Watanabe}}, \emph {et~al.},\ }\href@noop {} {\bibfield  {journal} {\bibinfo  {journal} {Nature}\ }\textbf {\bibinfo {volume} {650}},\ \bibinfo {pages} {580} (\bibinfo {year} {2026})}\BibitemShut {NoStop}%
	\bibitem [{\citenamefont {Huber}\ \emph {et~al.}(2026)\citenamefont {Huber}, \citenamefont {Kuhlbrodt}, \citenamefont {Anderson}, \citenamefont {Li}, \citenamefont {Watanabe}, \citenamefont {Taniguchi}, \citenamefont {Kroner}, \citenamefont {Xu}, \citenamefont {Imamo{\u{g}}lu},\ and\ \citenamefont {Smole{\'n}ski}}]{huber2026optical}%
	  \BibitemOpen
	  \bibfield  {author} {\bibinfo {author} {\bibfnamefont {O.}~\bibnamefont {Huber}}, \bibinfo {author} {\bibfnamefont {K.}~\bibnamefont {Kuhlbrodt}}, \bibinfo {author} {\bibfnamefont {E.}~\bibnamefont {Anderson}}, \bibinfo {author} {\bibfnamefont {W.}~\bibnamefont {Li}}, \bibinfo {author} {\bibfnamefont {K.}~\bibnamefont {Watanabe}}, \bibinfo {author} {\bibfnamefont {T.}~\bibnamefont {Taniguchi}}, \bibinfo {author} {\bibfnamefont {M.}~\bibnamefont {Kroner}}, \bibinfo {author} {\bibfnamefont {X.}~\bibnamefont {Xu}}, \bibinfo {author} {\bibfnamefont {A.}~\bibnamefont {Imamo{\u{g}}lu}},\ and\ \bibinfo {author} {\bibfnamefont {T.}~\bibnamefont {Smole{\'n}ski}},\ }\href@noop {} {\bibfield  {journal} {\bibinfo  {journal} {Nature}\ }\textbf {\bibinfo {volume} {649}},\ \bibinfo {pages} {1153} (\bibinfo {year} {2026})}\BibitemShut {NoStop}%
	\bibitem [{\citenamefont {Holtzmann}\ \emph {et~al.}(2026)\citenamefont {Holtzmann}, \citenamefont {Li}, \citenamefont {Anderson}, \citenamefont {Cai}, \citenamefont {Park}, \citenamefont {Hu}, \citenamefont {Taniguchi}, \citenamefont {Watanabe}, \citenamefont {Chu}, \citenamefont {Xiao} \emph {et~al.}}]{holtzmann2026optical}%
	  \BibitemOpen
	  \bibfield  {author} {\bibinfo {author} {\bibfnamefont {W.}~\bibnamefont {Holtzmann}}, \bibinfo {author} {\bibfnamefont {W.}~\bibnamefont {Li}}, \bibinfo {author} {\bibfnamefont {E.}~\bibnamefont {Anderson}}, \bibinfo {author} {\bibfnamefont {J.}~\bibnamefont {Cai}}, \bibinfo {author} {\bibfnamefont {H.}~\bibnamefont {Park}}, \bibinfo {author} {\bibfnamefont {C.}~\bibnamefont {Hu}}, \bibinfo {author} {\bibfnamefont {T.}~\bibnamefont {Taniguchi}}, \bibinfo {author} {\bibfnamefont {K.}~\bibnamefont {Watanabe}}, \bibinfo {author} {\bibfnamefont {J.-H.}\ \bibnamefont {Chu}}, \bibinfo {author} {\bibfnamefont {D.}~\bibnamefont {Xiao}}, \emph {et~al.},\ }\href@noop {} {\bibfield  {journal} {\bibinfo  {journal} {Nature}\ }\textbf {\bibinfo {volume} {649}},\ \bibinfo {pages} {1147} (\bibinfo {year} {2026})}\BibitemShut {NoStop}%
	\bibitem [{\citenamefont {Zhang}\ \emph {et~al.}(2024{\natexlab{a}})\citenamefont {Zhang}, \citenamefont {Wang}, \citenamefont {Liu}, \citenamefont {Fan}, \citenamefont {Cao},\ and\ \citenamefont {Xiao}}]{zhang2024polarization}%
	  \BibitemOpen
	  \bibfield  {author} {\bibinfo {author} {\bibfnamefont {X.-W.}\ \bibnamefont {Zhang}}, \bibinfo {author} {\bibfnamefont {C.}~\bibnamefont {Wang}}, \bibinfo {author} {\bibfnamefont {X.}~\bibnamefont {Liu}}, \bibinfo {author} {\bibfnamefont {Y.}~\bibnamefont {Fan}}, \bibinfo {author} {\bibfnamefont {T.}~\bibnamefont {Cao}},\ and\ \bibinfo {author} {\bibfnamefont {D.}~\bibnamefont {Xiao}},\ }\href@noop {} {\bibfield  {journal} {\bibinfo  {journal} {Nature Communications}\ }\textbf {\bibinfo {volume} {15}},\ \bibinfo {pages} {4223} (\bibinfo {year} {2024}{\natexlab{a}})}\BibitemShut {NoStop}%
	\bibitem [{\citenamefont {Morales-Dur{\'a}n}\ \emph {et~al.}(2023)\citenamefont {Morales-Dur{\'a}n}, \citenamefont {Wang}, \citenamefont {Schleder}, \citenamefont {Angeli}, \citenamefont {Zhu}, \citenamefont {Kaxiras}, \citenamefont {Repellin},\ and\ \citenamefont {Cano}}]{morales2023pressure}%
	  \BibitemOpen
	  \bibfield  {author} {\bibinfo {author} {\bibfnamefont {N.}~\bibnamefont {Morales-Dur{\'a}n}}, \bibinfo {author} {\bibfnamefont {J.}~\bibnamefont {Wang}}, \bibinfo {author} {\bibfnamefont {G.~R.}\ \bibnamefont {Schleder}}, \bibinfo {author} {\bibfnamefont {M.}~\bibnamefont {Angeli}}, \bibinfo {author} {\bibfnamefont {Z.}~\bibnamefont {Zhu}}, \bibinfo {author} {\bibfnamefont {E.}~\bibnamefont {Kaxiras}}, \bibinfo {author} {\bibfnamefont {C.}~\bibnamefont {Repellin}},\ and\ \bibinfo {author} {\bibfnamefont {J.}~\bibnamefont {Cano}},\ }\href@noop {} {\bibfield  {journal} {\bibinfo  {journal} {Physical Review Research}\ }\textbf {\bibinfo {volume} {5}},\ \bibinfo {pages} {L032022} (\bibinfo {year} {2023})}\BibitemShut {NoStop}%
	\bibitem [{\citenamefont {Jiao}\ \emph {et~al.}(2026)\citenamefont {Jiao}, \citenamefont {Qian}, \citenamefont {Mao}, \citenamefont {Chang}, \citenamefont {Xiao}, \citenamefont {Liu}, \citenamefont {Wang}, \citenamefont {Wu}, \citenamefont {Peng}, \citenamefont {Xu} \emph {et~al.}}]{jiao2026hydrostatic}%
	  \BibitemOpen
	  \bibfield  {author} {\bibinfo {author} {\bibfnamefont {P.}~\bibnamefont {Jiao}}, \bibinfo {author} {\bibfnamefont {C.}~\bibnamefont {Qian}}, \bibinfo {author} {\bibfnamefont {N.}~\bibnamefont {Mao}}, \bibinfo {author} {\bibfnamefont {X.}~\bibnamefont {Chang}}, \bibinfo {author} {\bibfnamefont {J.}~\bibnamefont {Xiao}}, \bibinfo {author} {\bibfnamefont {F.}~\bibnamefont {Liu}}, \bibinfo {author} {\bibfnamefont {S.}~\bibnamefont {Wang}}, \bibinfo {author} {\bibfnamefont {X.}~\bibnamefont {Wu}}, \bibinfo {author} {\bibfnamefont {D.}~\bibnamefont {Peng}}, \bibinfo {author} {\bibfnamefont {C.}~\bibnamefont {Xu}}, \emph {et~al.},\ }\href@noop {} {\bibfield  {journal} {\bibinfo  {journal} {Physical Review X}\ }\textbf {\bibinfo {volume} {16}},\ \bibinfo {pages} {011068} (\bibinfo {year} {2026})}\BibitemShut {NoStop}%
	\bibitem [{\citenamefont {Ding}\ \emph {et~al.}(2026)\citenamefont {Ding}, \citenamefont {Liang}, \citenamefont {Wu}, \citenamefont {Wu}, \citenamefont {L{\"u}}, \citenamefont {Gao},\ and\ \citenamefont {Xie}}]{ding2026sliding}%
	  \BibitemOpen
	  \bibfield  {author} {\bibinfo {author} {\bibfnamefont {S.-P.}\ \bibnamefont {Ding}}, \bibinfo {author} {\bibfnamefont {M.}~\bibnamefont {Liang}}, \bibinfo {author} {\bibfnamefont {T.-L.}\ \bibnamefont {Wu}}, \bibinfo {author} {\bibfnamefont {M.-H.}\ \bibnamefont {Wu}}, \bibinfo {author} {\bibfnamefont {J.-T.}\ \bibnamefont {L{\"u}}}, \bibinfo {author} {\bibfnamefont {J.-H.}\ \bibnamefont {Gao}},\ and\ \bibinfo {author} {\bibfnamefont {X.}~\bibnamefont {Xie}},\ }\href@noop {} {\bibfield  {journal} {\bibinfo  {journal} {Physical Review B}\ }\textbf {\bibinfo {volume} {113}},\ \bibinfo {pages} {L121411} (\bibinfo {year} {2026})}\BibitemShut {NoStop}%
	\bibitem [{\citenamefont {Zheng}\ \emph {et~al.}(2024)\citenamefont {Zheng}, \citenamefont {Zhai}, \citenamefont {Xiao},\ and\ \citenamefont {Yao}}]{zheng2024interlayer}%
	  \BibitemOpen
	  \bibfield  {author} {\bibinfo {author} {\bibfnamefont {H.}~\bibnamefont {Zheng}}, \bibinfo {author} {\bibfnamefont {D.}~\bibnamefont {Zhai}}, \bibinfo {author} {\bibfnamefont {C.}~\bibnamefont {Xiao}},\ and\ \bibinfo {author} {\bibfnamefont {W.}~\bibnamefont {Yao}},\ }\href@noop {} {\bibfield  {journal} {\bibinfo  {journal} {Nano Letters}\ }\textbf {\bibinfo {volume} {24}},\ \bibinfo {pages} {8017} (\bibinfo {year} {2024})}\BibitemShut {NoStop}%
	\bibitem [{\citenamefont {Liang}\ \emph {et~al.}(2025)\citenamefont {Liang}, \citenamefont {Ding}, \citenamefont {Wu}, \citenamefont {Zhao},\ and\ \citenamefont {Gao}}]{liang2025moire}%
	  \BibitemOpen
	  \bibfield  {author} {\bibinfo {author} {\bibfnamefont {M.}~\bibnamefont {Liang}}, \bibinfo {author} {\bibfnamefont {S.-P.}\ \bibnamefont {Ding}}, \bibinfo {author} {\bibfnamefont {M.}~\bibnamefont {Wu}}, \bibinfo {author} {\bibfnamefont {C.}~\bibnamefont {Zhao}},\ and\ \bibinfo {author} {\bibfnamefont {J.-H.}\ \bibnamefont {Gao}},\ }\href@noop {} {\bibfield  {journal} {\bibinfo  {journal} {Physical Review B}\ }\textbf {\bibinfo {volume} {111}},\ \bibinfo {pages} {085412} (\bibinfo {year} {2025})}\BibitemShut {NoStop}%
	\bibitem [{\citenamefont {Nakatsuji}\ \emph {et~al.}(2025)\citenamefont {Nakatsuji}, \citenamefont {Kawakami}, \citenamefont {Tateishi}, \citenamefont {Kato},\ and\ \citenamefont {Koshino}}]{nakatsuji2025moire}%
	  \BibitemOpen
	  \bibfield  {author} {\bibinfo {author} {\bibfnamefont {N.}~\bibnamefont {Nakatsuji}}, \bibinfo {author} {\bibfnamefont {T.}~\bibnamefont {Kawakami}}, \bibinfo {author} {\bibfnamefont {H.}~\bibnamefont {Tateishi}}, \bibinfo {author} {\bibfnamefont {K.}~\bibnamefont {Kato}},\ and\ \bibinfo {author} {\bibfnamefont {M.}~\bibnamefont {Koshino}},\ }\href@noop {} {\bibfield  {journal} {\bibinfo  {journal} {Communications Materials}\ }\textbf {\bibinfo {volume} {6}},\ \bibinfo {pages} {274} (\bibinfo {year} {2025})}\BibitemShut {NoStop}%
	\bibitem [{\citenamefont {Choi}\ \emph {et~al.}(2025)\citenamefont {Choi}, \citenamefont {Morales-Dur{\'a}n}, \citenamefont {Kwan}, \citenamefont {Millis}, \citenamefont {Regnault},\ and\ \citenamefont {Guerci}}]{choi2025higher}%
	  \BibitemOpen
	  \bibfield  {author} {\bibinfo {author} {\bibfnamefont {J.~D.}\ \bibnamefont {Choi}}, \bibinfo {author} {\bibfnamefont {N.}~\bibnamefont {Morales-Dur{\'a}n}}, \bibinfo {author} {\bibfnamefont {Y.~H.}\ \bibnamefont {Kwan}}, \bibinfo {author} {\bibfnamefont {A.~J.}\ \bibnamefont {Millis}}, \bibinfo {author} {\bibfnamefont {N.}~\bibnamefont {Regnault}},\ and\ \bibinfo {author} {\bibfnamefont {D.}~\bibnamefont {Guerci}},\ }\href@noop {} {\bibfield  {journal} {\bibinfo  {journal} {Physical Review B}\ }\textbf {\bibinfo {volume} {112}},\ \bibinfo {pages} {205122} (\bibinfo {year} {2025})}\BibitemShut {NoStop}%
	\bibitem [{\citenamefont {Fan}\ \emph {et~al.}(2026)\citenamefont {Fan}, \citenamefont {Zhang}, \citenamefont {Ye}, \citenamefont {Liu}, \citenamefont {Wang}, \citenamefont {Yang}, \citenamefont {Xiao},\ and\ \citenamefont {Cao}}]{fan2026layerwise}%
	  \BibitemOpen
	  \bibfield  {author} {\bibinfo {author} {\bibfnamefont {Y.}~\bibnamefont {Fan}}, \bibinfo {author} {\bibfnamefont {X.-W.}\ \bibnamefont {Zhang}}, \bibinfo {author} {\bibfnamefont {Y.}~\bibnamefont {Ye}}, \bibinfo {author} {\bibfnamefont {X.}~\bibnamefont {Liu}}, \bibinfo {author} {\bibfnamefont {C.}~\bibnamefont {Wang}}, \bibinfo {author} {\bibfnamefont {K.}~\bibnamefont {Yang}}, \bibinfo {author} {\bibfnamefont {D.}~\bibnamefont {Xiao}},\ and\ \bibinfo {author} {\bibfnamefont {T.}~\bibnamefont {Cao}},\ }\href@noop {} {\bibfield  {journal} {\bibinfo  {journal} {Proceedings of the National Academy of Sciences}\ }\textbf {\bibinfo {volume} {123}},\ \bibinfo {pages} {e2532550123} (\bibinfo {year} {2026})}\BibitemShut {NoStop}%
	\bibitem [{\citenamefont {Qi}\ \emph {et~al.}(2026)\citenamefont {Qi}, \citenamefont {Pi}, \citenamefont {Zhang}, \citenamefont {Liu}, \citenamefont {Regnault}, \citenamefont {Weng}, \citenamefont {Bernevig}, \citenamefont {Yu},\ and\ \citenamefont {Wu}}]{qi2026chern}%
	  \BibitemOpen
	  \bibfield  {author} {\bibinfo {author} {\bibfnamefont {Z.}~\bibnamefont {Qi}}, \bibinfo {author} {\bibfnamefont {H.}~\bibnamefont {Pi}}, \bibinfo {author} {\bibfnamefont {Y.}~\bibnamefont {Zhang}}, \bibinfo {author} {\bibfnamefont {J.}~\bibnamefont {Liu}}, \bibinfo {author} {\bibfnamefont {N.}~\bibnamefont {Regnault}}, \bibinfo {author} {\bibfnamefont {H.}~\bibnamefont {Weng}}, \bibinfo {author} {\bibfnamefont {B.~A.}\ \bibnamefont {Bernevig}}, \bibinfo {author} {\bibfnamefont {J.}~\bibnamefont {Yu}},\ and\ \bibinfo {author} {\bibfnamefont {Q.}~\bibnamefont {Wu}},\ }\href@noop {} {\bibfield  {journal} {\bibinfo  {journal} {Physical Review B}\ }\textbf {\bibinfo {volume} {113}},\ \bibinfo {pages} {125116} (\bibinfo {year} {2026})}\BibitemShut {NoStop}%
	\bibitem [{\citenamefont {Li}\ \emph {et~al.}(2026)\citenamefont {Li}, \citenamefont {Qiu}, \citenamefont {Wu},\ and\ \citenamefont {MacDonald}}]{li2026quantum}%
	  \BibitemOpen
	  \bibfield  {author} {\bibinfo {author} {\bibfnamefont {B.}~\bibnamefont {Li}}, \bibinfo {author} {\bibfnamefont {W.-X.}\ \bibnamefont {Qiu}}, \bibinfo {author} {\bibfnamefont {F.}~\bibnamefont {Wu}},\ and\ \bibinfo {author} {\bibfnamefont {A.}~\bibnamefont {MacDonald}},\ }\href@noop {} {\bibfield  {journal} {\bibinfo  {journal} {National Science Review}\ }\textbf {\bibinfo {volume} {13}},\ \bibinfo {pages} {nwaf570} (\bibinfo {year} {2026})}\BibitemShut {NoStop}%
	\bibitem [{\citenamefont {Jia}\ \emph {et~al.}(2024)\citenamefont {Jia}, \citenamefont {Yu}, \citenamefont {Liu}, \citenamefont {Herzog-Arbeitman}, \citenamefont {Qi}, \citenamefont {Pi}, \citenamefont {Regnault}, \citenamefont {Weng}, \citenamefont {Bernevig},\ and\ \citenamefont {Wu}}]{jia2024moire}%
	  \BibitemOpen
	  \bibfield  {author} {\bibinfo {author} {\bibfnamefont {Y.}~\bibnamefont {Jia}}, \bibinfo {author} {\bibfnamefont {J.}~\bibnamefont {Yu}}, \bibinfo {author} {\bibfnamefont {J.}~\bibnamefont {Liu}}, \bibinfo {author} {\bibfnamefont {J.}~\bibnamefont {Herzog-Arbeitman}}, \bibinfo {author} {\bibfnamefont {Z.}~\bibnamefont {Qi}}, \bibinfo {author} {\bibfnamefont {H.}~\bibnamefont {Pi}}, \bibinfo {author} {\bibfnamefont {N.}~\bibnamefont {Regnault}}, \bibinfo {author} {\bibfnamefont {H.}~\bibnamefont {Weng}}, \bibinfo {author} {\bibfnamefont {B.~A.}\ \bibnamefont {Bernevig}},\ and\ \bibinfo {author} {\bibfnamefont {Q.}~\bibnamefont {Wu}},\ }\href@noop {} {\bibfield  {journal} {\bibinfo  {journal} {Physical Review B}\ }\textbf {\bibinfo {volume} {109}},\ \bibinfo {pages} {205121} (\bibinfo {year} {2024})}\BibitemShut {NoStop}%
	\bibitem [{\citenamefont {Wang}\ \emph {et~al.}(2018)\citenamefont {Wang}, \citenamefont {Zhang}, \citenamefont {Han} \emph {et~al.}}]{wang2018deepmd}%
	  \BibitemOpen
	  \bibfield  {author} {\bibinfo {author} {\bibfnamefont {H.}~\bibnamefont {Wang}}, \bibinfo {author} {\bibfnamefont {L.}~\bibnamefont {Zhang}}, \bibinfo {author} {\bibfnamefont {J.}~\bibnamefont {Han}}, \emph {et~al.},\ }\href@noop {} {\bibfield  {journal} {\bibinfo  {journal} {Computer Physics Communications}\ }\textbf {\bibinfo {volume} {228}},\ \bibinfo {pages} {178} (\bibinfo {year} {2018})}\BibitemShut {NoStop}%
	\bibitem [{\citenamefont {Zeng}\ \emph {et~al.}(2023{\natexlab{b}})\citenamefont {Zeng}, \citenamefont {Zhang}, \citenamefont {Lu}, \citenamefont {Mo}, \citenamefont {Li}, \citenamefont {Chen}, \citenamefont {Rynik}, \citenamefont {Huang}, \citenamefont {Li}, \citenamefont {Shi} \emph {et~al.}}]{zeng2023deepmd}%
	  \BibitemOpen
	  \bibfield  {author} {\bibinfo {author} {\bibfnamefont {J.}~\bibnamefont {Zeng}}, \bibinfo {author} {\bibfnamefont {D.}~\bibnamefont {Zhang}}, \bibinfo {author} {\bibfnamefont {D.}~\bibnamefont {Lu}}, \bibinfo {author} {\bibfnamefont {P.}~\bibnamefont {Mo}}, \bibinfo {author} {\bibfnamefont {Z.}~\bibnamefont {Li}}, \bibinfo {author} {\bibfnamefont {Y.}~\bibnamefont {Chen}}, \bibinfo {author} {\bibfnamefont {M.}~\bibnamefont {Rynik}}, \bibinfo {author} {\bibfnamefont {L.}~\bibnamefont {Huang}}, \bibinfo {author} {\bibfnamefont {Z.}~\bibnamefont {Li}}, \bibinfo {author} {\bibfnamefont {S.}~\bibnamefont {Shi}}, \emph {et~al.},\ }\href@noop {} {\bibfield  {journal} {\bibinfo  {journal} {The Journal of Chemical Physics}\ }\textbf {\bibinfo {volume} {159}} (\bibinfo {year} {2023}{\natexlab{b}})}\BibitemShut {NoStop}%
	\bibitem [{\citenamefont {Kresse}\ and\ \citenamefont {Furthm{\"u}ller}(1996)}]{kresse1996efficiency}%
	  \BibitemOpen
	  \bibfield  {author} {\bibinfo {author} {\bibfnamefont {G.}~\bibnamefont {Kresse}}\ and\ \bibinfo {author} {\bibfnamefont {J.}~\bibnamefont {Furthm{\"u}ller}},\ }\href@noop {} {\bibfield  {journal} {\bibinfo  {journal} {Computational materials science}\ }\textbf {\bibinfo {volume} {6}},\ \bibinfo {pages} {15} (\bibinfo {year} {1996})}\BibitemShut {NoStop}%
	\bibitem [{sup()}]{supp}%
	  \BibitemOpen
	  \href@noop {} {\bibinfo  {journal} {See Supplemental Materials for the numerical details of parameterizations of MLFFs, lattice relaxations, DFT band calculations, derivations and fitting of continuum model, and moir\'e potential distributions. Refs.~\cite{blochl1994projector,kresse1999ultrasoft,kresse1996efficiency,perdew1996generalized,grimme2006semiempirical,wang2018deepmd,zeng2023deepmd,thompson2022lammps,agarwal1972measurement,soler2002siesta,hamann2013optimized,fernandez2006site,zhang2024polarization,zhang2025twist} are included.}\ }\BibitemShut {NoStop}%
	\bibitem [{\citenamefont {Soler}\ \emph {et~al.}(2002)\citenamefont {Soler}, \citenamefont {Artacho}, \citenamefont {Gale}, \citenamefont {Garc{\'\i}a}, \citenamefont {Junquera}, \citenamefont {Ordej{\'o}n},\ and\ \citenamefont {S{\'a}nchez-Portal}}]{soler2002siesta}%
	  \BibitemOpen
	\bibfield  {journal} {  }\bibfield  {author} {\bibinfo {author} {\bibfnamefont {J.~M.}\ \bibnamefont {Soler}}, \bibinfo {author} {\bibfnamefont {E.}~\bibnamefont {Artacho}}, \bibinfo {author} {\bibfnamefont {J.~D.}\ \bibnamefont {Gale}}, \bibinfo {author} {\bibfnamefont {A.}~\bibnamefont {Garc{\'\i}a}}, \bibinfo {author} {\bibfnamefont {J.}~\bibnamefont {Junquera}}, \bibinfo {author} {\bibfnamefont {P.}~\bibnamefont {Ordej{\'o}n}},\ and\ \bibinfo {author} {\bibfnamefont {D.}~\bibnamefont {S{\'a}nchez-Portal}},\ }\href@noop {} {\bibfield  {journal} {\bibinfo  {journal} {Journal of physics: Condensed matter}\ }\textbf {\bibinfo {volume} {14}},\ \bibinfo {pages} {2745} (\bibinfo {year} {2002})}\BibitemShut {NoStop}%
	\bibitem [{\citenamefont {Wang}\ \emph {et~al.}(2024)\citenamefont {Wang}, \citenamefont {Zhang}, \citenamefont {Liu}, \citenamefont {He}, \citenamefont {Xu}, \citenamefont {Ran}, \citenamefont {Cao},\ and\ \citenamefont {Xiao}}]{wang2024fractional}%
	  \BibitemOpen
	  \bibfield  {author} {\bibinfo {author} {\bibfnamefont {C.}~\bibnamefont {Wang}}, \bibinfo {author} {\bibfnamefont {X.-W.}\ \bibnamefont {Zhang}}, \bibinfo {author} {\bibfnamefont {X.}~\bibnamefont {Liu}}, \bibinfo {author} {\bibfnamefont {Y.}~\bibnamefont {He}}, \bibinfo {author} {\bibfnamefont {X.}~\bibnamefont {Xu}}, \bibinfo {author} {\bibfnamefont {Y.}~\bibnamefont {Ran}}, \bibinfo {author} {\bibfnamefont {T.}~\bibnamefont {Cao}},\ and\ \bibinfo {author} {\bibfnamefont {D.}~\bibnamefont {Xiao}},\ }\href@noop {} {\bibfield  {journal} {\bibinfo  {journal} {Physical Review Letters}\ }\textbf {\bibinfo {volume} {132}},\ \bibinfo {pages} {036501} (\bibinfo {year} {2024})}\BibitemShut {NoStop}%
	\bibitem [{\citenamefont {Zhang}\ \emph {et~al.}(2025)\citenamefont {Zhang}, \citenamefont {Yang}, \citenamefont {Wang}, \citenamefont {Liu}, \citenamefont {Cao},\ and\ \citenamefont {Xiao}}]{zhang2025twist}%
	  \BibitemOpen
	  \bibfield  {author} {\bibinfo {author} {\bibfnamefont {X.-W.}\ \bibnamefont {Zhang}}, \bibinfo {author} {\bibfnamefont {K.}~\bibnamefont {Yang}}, \bibinfo {author} {\bibfnamefont {C.}~\bibnamefont {Wang}}, \bibinfo {author} {\bibfnamefont {X.}~\bibnamefont {Liu}}, \bibinfo {author} {\bibfnamefont {T.}~\bibnamefont {Cao}},\ and\ \bibinfo {author} {\bibfnamefont {D.}~\bibnamefont {Xiao}},\ }\href@noop {} {\bibfield  {journal} {\bibinfo  {journal} {npj Quantum Materials}\ }\textbf {\bibinfo {volume} {10}},\ \bibinfo {pages} {110} (\bibinfo {year} {2025})}\BibitemShut {NoStop}%
	\bibitem [{\citenamefont {Shi}\ \emph {et~al.}(2021)\citenamefont {Shi}, \citenamefont {Zhu},\ and\ \citenamefont {MacDonald}}]{shi2021moire}%
	  \BibitemOpen
	  \bibfield  {author} {\bibinfo {author} {\bibfnamefont {J.}~\bibnamefont {Shi}}, \bibinfo {author} {\bibfnamefont {J.}~\bibnamefont {Zhu}},\ and\ \bibinfo {author} {\bibfnamefont {A.}~\bibnamefont {MacDonald}},\ }\href@noop {} {\bibfield  {journal} {\bibinfo  {journal} {Physical Review B}\ }\textbf {\bibinfo {volume} {103}},\ \bibinfo {pages} {075122} (\bibinfo {year} {2021})}\BibitemShut {NoStop}%
	\bibitem [{\citenamefont {Grover}\ \emph {et~al.}(2022)\citenamefont {Grover}, \citenamefont {Bocarsly}, \citenamefont {Uri}, \citenamefont {Stepanov}, \citenamefont {Di~Battista}, \citenamefont {Roy}, \citenamefont {Xiao}, \citenamefont {Meltzer}, \citenamefont {Myasoedov}, \citenamefont {Pareek} \emph {et~al.}}]{grover2022chern}%
	  \BibitemOpen
	  \bibfield  {author} {\bibinfo {author} {\bibfnamefont {S.}~\bibnamefont {Grover}}, \bibinfo {author} {\bibfnamefont {M.}~\bibnamefont {Bocarsly}}, \bibinfo {author} {\bibfnamefont {A.}~\bibnamefont {Uri}}, \bibinfo {author} {\bibfnamefont {P.}~\bibnamefont {Stepanov}}, \bibinfo {author} {\bibfnamefont {G.}~\bibnamefont {Di~Battista}}, \bibinfo {author} {\bibfnamefont {I.}~\bibnamefont {Roy}}, \bibinfo {author} {\bibfnamefont {J.}~\bibnamefont {Xiao}}, \bibinfo {author} {\bibfnamefont {A.~Y.}\ \bibnamefont {Meltzer}}, \bibinfo {author} {\bibfnamefont {Y.}~\bibnamefont {Myasoedov}}, \bibinfo {author} {\bibfnamefont {K.}~\bibnamefont {Pareek}}, \emph {et~al.},\ }\href@noop {} {\bibfield  {journal} {\bibinfo  {journal} {Nature physics}\ }\textbf {\bibinfo {volume} {18}},\ \bibinfo {pages} {885} (\bibinfo {year} {2022})}\BibitemShut {NoStop}%
	\bibitem [{\citenamefont {Devakul}\ \emph {et~al.}(2023)\citenamefont {Devakul}, \citenamefont {Ledwith}, \citenamefont {Xia}, \citenamefont {Uri}, \citenamefont {de~la Barrera}, \citenamefont {Jarillo-Herrero},\ and\ \citenamefont {Fu}}]{devakul2023magic}%
	  \BibitemOpen
	  \bibfield  {author} {\bibinfo {author} {\bibfnamefont {T.}~\bibnamefont {Devakul}}, \bibinfo {author} {\bibfnamefont {P.~J.}\ \bibnamefont {Ledwith}}, \bibinfo {author} {\bibfnamefont {L.-Q.}\ \bibnamefont {Xia}}, \bibinfo {author} {\bibfnamefont {A.}~\bibnamefont {Uri}}, \bibinfo {author} {\bibfnamefont {S.~C.}\ \bibnamefont {de~la Barrera}}, \bibinfo {author} {\bibfnamefont {P.}~\bibnamefont {Jarillo-Herrero}},\ and\ \bibinfo {author} {\bibfnamefont {L.}~\bibnamefont {Fu}},\ }\href@noop {} {\bibfield  {journal} {\bibinfo  {journal} {Science Advances}\ }\textbf {\bibinfo {volume} {9}},\ \bibinfo {pages} {eadi6063} (\bibinfo {year} {2023})}\BibitemShut {NoStop}%
	\bibitem [{\citenamefont {Zhang}\ \emph {et~al.}(2024{\natexlab{b}})\citenamefont {Zhang}, \citenamefont {Zhu}, \citenamefont {Kahn}, \citenamefont {Soejima}, \citenamefont {Watanabe}, \citenamefont {Taniguchi}, \citenamefont {Zettl}, \citenamefont {Wang}, \citenamefont {Zaletel},\ and\ \citenamefont {Crommie}}]{zhang2024manipulation}%
	  \BibitemOpen
	  \bibfield  {author} {\bibinfo {author} {\bibfnamefont {C.}~\bibnamefont {Zhang}}, \bibinfo {author} {\bibfnamefont {T.}~\bibnamefont {Zhu}}, \bibinfo {author} {\bibfnamefont {S.}~\bibnamefont {Kahn}}, \bibinfo {author} {\bibfnamefont {T.}~\bibnamefont {Soejima}}, \bibinfo {author} {\bibfnamefont {K.}~\bibnamefont {Watanabe}}, \bibinfo {author} {\bibfnamefont {T.}~\bibnamefont {Taniguchi}}, \bibinfo {author} {\bibfnamefont {A.}~\bibnamefont {Zettl}}, \bibinfo {author} {\bibfnamefont {F.}~\bibnamefont {Wang}}, \bibinfo {author} {\bibfnamefont {M.~P.}\ \bibnamefont {Zaletel}},\ and\ \bibinfo {author} {\bibfnamefont {M.~F.}\ \bibnamefont {Crommie}},\ }\href@noop {} {\bibfield  {journal} {\bibinfo  {journal} {Nature Physics}\ }\textbf {\bibinfo {volume} {20}},\ \bibinfo {pages} {951} (\bibinfo {year} {2024}{\natexlab{b}})}\BibitemShut {NoStop}%
	\bibitem [{\citenamefont {Xia}\ \emph {et~al.}(2025)\citenamefont {Xia}, \citenamefont {de~la Barrera}, \citenamefont {Uri}, \citenamefont {Sharpe}, \citenamefont {Kwan}, \citenamefont {Zhu}, \citenamefont {Watanabe}, \citenamefont {Taniguchi}, \citenamefont {Goldhaber-Gordon}, \citenamefont {Fu} \emph {et~al.}}]{xia2025topological}%
	  \BibitemOpen
	  \bibfield  {author} {\bibinfo {author} {\bibfnamefont {L.-Q.}\ \bibnamefont {Xia}}, \bibinfo {author} {\bibfnamefont {S.~C.}\ \bibnamefont {de~la Barrera}}, \bibinfo {author} {\bibfnamefont {A.}~\bibnamefont {Uri}}, \bibinfo {author} {\bibfnamefont {A.}~\bibnamefont {Sharpe}}, \bibinfo {author} {\bibfnamefont {Y.~H.}\ \bibnamefont {Kwan}}, \bibinfo {author} {\bibfnamefont {Z.}~\bibnamefont {Zhu}}, \bibinfo {author} {\bibfnamefont {K.}~\bibnamefont {Watanabe}}, \bibinfo {author} {\bibfnamefont {T.}~\bibnamefont {Taniguchi}}, \bibinfo {author} {\bibfnamefont {D.}~\bibnamefont {Goldhaber-Gordon}}, \bibinfo {author} {\bibfnamefont {L.}~\bibnamefont {Fu}}, \emph {et~al.},\ }\href@noop {} {\bibfield  {journal} {\bibinfo  {journal} {Nature Physics}\ }\textbf {\bibinfo {volume} {21}},\ \bibinfo {pages} {239} (\bibinfo {year} {2025})}\BibitemShut {NoStop}%
	\bibitem [{\citenamefont {Fujimoto}\ \emph {et~al.}(2025)\citenamefont {Fujimoto}, \citenamefont {Nakatsuji}, \citenamefont {Vishwanath},\ and\ \citenamefont {Ledwith}}]{fujimoto2025four}%
	  \BibitemOpen
	  \bibfield  {author} {\bibinfo {author} {\bibfnamefont {M.}~\bibnamefont {Fujimoto}}, \bibinfo {author} {\bibfnamefont {N.}~\bibnamefont {Nakatsuji}}, \bibinfo {author} {\bibfnamefont {A.}~\bibnamefont {Vishwanath}},\ and\ \bibinfo {author} {\bibfnamefont {P.}~\bibnamefont {Ledwith}},\ }\href@noop {} {\bibfield  {journal} {\bibinfo  {journal} {arXiv preprint arXiv:2510.02444}\ } (\bibinfo {year} {2025})}\BibitemShut {NoStop}%
	\bibitem [{\citenamefont {Tong}\ \emph {et~al.}(2017)\citenamefont {Tong}, \citenamefont {Yu}, \citenamefont {Zhu}, \citenamefont {Wang}, \citenamefont {Xu},\ and\ \citenamefont {Yao}}]{tong2017topological}%
	  \BibitemOpen
	  \bibfield  {author} {\bibinfo {author} {\bibfnamefont {Q.}~\bibnamefont {Tong}}, \bibinfo {author} {\bibfnamefont {H.}~\bibnamefont {Yu}}, \bibinfo {author} {\bibfnamefont {Q.}~\bibnamefont {Zhu}}, \bibinfo {author} {\bibfnamefont {Y.}~\bibnamefont {Wang}}, \bibinfo {author} {\bibfnamefont {X.}~\bibnamefont {Xu}},\ and\ \bibinfo {author} {\bibfnamefont {W.}~\bibnamefont {Yao}},\ }\href@noop {} {\bibfield  {journal} {\bibinfo  {journal} {Nature Physics}\ }\textbf {\bibinfo {volume} {13}},\ \bibinfo {pages} {356} (\bibinfo {year} {2017})}\BibitemShut {NoStop}%
	\bibitem [{\citenamefont {Ovchinnikov}\ \emph {et~al.}(2022)\citenamefont {Ovchinnikov}, \citenamefont {Cai}, \citenamefont {Lin}, \citenamefont {Fei}, \citenamefont {Liu}, \citenamefont {Cui}, \citenamefont {Cobden}, \citenamefont {Chu}, \citenamefont {Chang}, \citenamefont {Xiao} \emph {et~al.}}]{ovchinnikov2022topological}%
	  \BibitemOpen
	  \bibfield  {author} {\bibinfo {author} {\bibfnamefont {D.}~\bibnamefont {Ovchinnikov}}, \bibinfo {author} {\bibfnamefont {J.}~\bibnamefont {Cai}}, \bibinfo {author} {\bibfnamefont {Z.}~\bibnamefont {Lin}}, \bibinfo {author} {\bibfnamefont {Z.}~\bibnamefont {Fei}}, \bibinfo {author} {\bibfnamefont {Z.}~\bibnamefont {Liu}}, \bibinfo {author} {\bibfnamefont {Y.-T.}\ \bibnamefont {Cui}}, \bibinfo {author} {\bibfnamefont {D.~H.}\ \bibnamefont {Cobden}}, \bibinfo {author} {\bibfnamefont {J.-H.}\ \bibnamefont {Chu}}, \bibinfo {author} {\bibfnamefont {C.-Z.}\ \bibnamefont {Chang}}, \bibinfo {author} {\bibfnamefont {D.}~\bibnamefont {Xiao}}, \emph {et~al.},\ }\href@noop {} {\bibfield  {journal} {\bibinfo  {journal} {Nature Communications}\ }\textbf {\bibinfo {volume} {13}},\ \bibinfo {pages} {5967} (\bibinfo {year} {2022})}\BibitemShut {NoStop}%
	\bibitem [{\citenamefont {Zhao}\ \emph {et~al.}(2023)\citenamefont {Zhao}, \citenamefont {Zhang}, \citenamefont {Cai}, \citenamefont {Zhuo}, \citenamefont {Zhou}, \citenamefont {Yan}, \citenamefont {Chan}, \citenamefont {Xu},\ and\ \citenamefont {Chang}}]{zhao2023creation}%
	  \BibitemOpen
	  \bibfield  {author} {\bibinfo {author} {\bibfnamefont {Y.-F.}\ \bibnamefont {Zhao}}, \bibinfo {author} {\bibfnamefont {R.}~\bibnamefont {Zhang}}, \bibinfo {author} {\bibfnamefont {J.}~\bibnamefont {Cai}}, \bibinfo {author} {\bibfnamefont {D.}~\bibnamefont {Zhuo}}, \bibinfo {author} {\bibfnamefont {L.-J.}\ \bibnamefont {Zhou}}, \bibinfo {author} {\bibfnamefont {Z.-J.}\ \bibnamefont {Yan}}, \bibinfo {author} {\bibfnamefont {M.~H.}\ \bibnamefont {Chan}}, \bibinfo {author} {\bibfnamefont {X.}~\bibnamefont {Xu}},\ and\ \bibinfo {author} {\bibfnamefont {C.-Z.}\ \bibnamefont {Chang}},\ }\href@noop {} {\bibfield  {journal} {\bibinfo  {journal} {Nature Communications}\ }\textbf {\bibinfo {volume} {14}},\ \bibinfo {pages} {770} (\bibinfo {year} {2023})}\BibitemShut {NoStop}%
	\bibitem [{\citenamefont {Yan}\ \emph {et~al.}(2024)\citenamefont {Yan}, \citenamefont {Li}, \citenamefont {Jiang}, \citenamefont {Sun},\ and\ \citenamefont {Xie}}]{yan2024rules}%
	  \BibitemOpen
	  \bibfield  {author} {\bibinfo {author} {\bibfnamefont {Q.}~\bibnamefont {Yan}}, \bibinfo {author} {\bibfnamefont {H.}~\bibnamefont {Li}}, \bibinfo {author} {\bibfnamefont {H.}~\bibnamefont {Jiang}}, \bibinfo {author} {\bibfnamefont {Q.-F.}\ \bibnamefont {Sun}},\ and\ \bibinfo {author} {\bibfnamefont {X.}~\bibnamefont {Xie}},\ }\href@noop {} {\bibfield  {journal} {\bibinfo  {journal} {Science Advances}\ }\textbf {\bibinfo {volume} {10}},\ \bibinfo {pages} {eado4756} (\bibinfo {year} {2024})}\BibitemShut {NoStop}%
	\bibitem [{\citenamefont {Gilbert}(2025)}]{gilbert2025chern}%
	  \BibitemOpen
	  \bibfield  {author} {\bibinfo {author} {\bibfnamefont {M.~J.}\ \bibnamefont {Gilbert}},\ }\href@noop {} {\bibfield  {journal} {\bibinfo  {journal} {Nature Communications}\ }\textbf {\bibinfo {volume} {16}},\ \bibinfo {pages} {3904} (\bibinfo {year} {2025})}\BibitemShut {NoStop}%
	\bibitem [{\citenamefont {Bhattacharjee}\ \emph {et~al.}(2026)\citenamefont {Bhattacharjee}, \citenamefont {May-Mann}, \citenamefont {Kwan}, \citenamefont {Devakul},\ and\ \citenamefont {Sharpe}}]{bhattacharjee2026mesoscopic}%
	  \BibitemOpen
	  \bibfield  {author} {\bibinfo {author} {\bibfnamefont {S.}~\bibnamefont {Bhattacharjee}}, \bibinfo {author} {\bibfnamefont {J.}~\bibnamefont {May-Mann}}, \bibinfo {author} {\bibfnamefont {Y.~H.}\ \bibnamefont {Kwan}}, \bibinfo {author} {\bibfnamefont {T.}~\bibnamefont {Devakul}},\ and\ \bibinfo {author} {\bibfnamefont {A.}~\bibnamefont {Sharpe}},\ }\href@noop {} {\bibfield  {journal} {\bibinfo  {journal} {arXiv preprint arXiv:2604.08654}\ } (\bibinfo {year} {2026})}\BibitemShut {NoStop}%
	\bibitem [{\citenamefont {Bl{\"o}chl}(1994)}]{blochl1994projector}%
	  \BibitemOpen
	  \bibfield  {author} {\bibinfo {author} {\bibfnamefont {P.~E.}\ \bibnamefont {Bl{\"o}chl}},\ }\href@noop {} {\bibfield  {journal} {\bibinfo  {journal} {Physical Review B}\ }\textbf {\bibinfo {volume} {50}},\ \bibinfo {pages} {17953} (\bibinfo {year} {1994})}\BibitemShut {NoStop}%
	\bibitem [{\citenamefont {Kresse}\ and\ \citenamefont {Joubert}(1999)}]{kresse1999ultrasoft}%
	  \BibitemOpen
	  \bibfield  {author} {\bibinfo {author} {\bibfnamefont {G.}~\bibnamefont {Kresse}}\ and\ \bibinfo {author} {\bibfnamefont {D.}~\bibnamefont {Joubert}},\ }\href@noop {} {\bibfield  {journal} {\bibinfo  {journal} {Physical Review B}\ }\textbf {\bibinfo {volume} {59}},\ \bibinfo {pages} {1758} (\bibinfo {year} {1999})}\BibitemShut {NoStop}%
	\bibitem [{\citenamefont {Perdew}\ \emph {et~al.}(1996)\citenamefont {Perdew}, \citenamefont {Burke},\ and\ \citenamefont {Ernzerhof}}]{perdew1996generalized}%
	  \BibitemOpen
	  \bibfield  {author} {\bibinfo {author} {\bibfnamefont {J.~P.}\ \bibnamefont {Perdew}}, \bibinfo {author} {\bibfnamefont {K.}~\bibnamefont {Burke}},\ and\ \bibinfo {author} {\bibfnamefont {M.}~\bibnamefont {Ernzerhof}},\ }\href@noop {} {\bibfield  {journal} {\bibinfo  {journal} {Physical Review Letters}\ }\textbf {\bibinfo {volume} {77}},\ \bibinfo {pages} {3865} (\bibinfo {year} {1996})}\BibitemShut {NoStop}%
	\bibitem [{\citenamefont {Grimme}(2006)}]{grimme2006semiempirical}%
	  \BibitemOpen
	  \bibfield  {author} {\bibinfo {author} {\bibfnamefont {S.}~\bibnamefont {Grimme}},\ }\href@noop {} {\bibfield  {journal} {\bibinfo  {journal} {Journal of Computational Chemistry}\ }\textbf {\bibinfo {volume} {27}},\ \bibinfo {pages} {1787} (\bibinfo {year} {2006})}\BibitemShut {NoStop}%
	\bibitem [{\citenamefont {Thompson}\ \emph {et~al.}(2022)\citenamefont {Thompson}, \citenamefont {Aktulga}, \citenamefont {Berger}, \citenamefont {Bolintineanu}, \citenamefont {Brown}, \citenamefont {Crozier}, \citenamefont {In't~Veld}, \citenamefont {Kohlmeyer}, \citenamefont {Moore}, \citenamefont {Nguyen} \emph {et~al.}}]{thompson2022lammps}%
	  \BibitemOpen
	  \bibfield  {author} {\bibinfo {author} {\bibfnamefont {A.~P.}\ \bibnamefont {Thompson}}, \bibinfo {author} {\bibfnamefont {H.~M.}\ \bibnamefont {Aktulga}}, \bibinfo {author} {\bibfnamefont {R.}~\bibnamefont {Berger}}, \bibinfo {author} {\bibfnamefont {D.~S.}\ \bibnamefont {Bolintineanu}}, \bibinfo {author} {\bibfnamefont {W.~M.}\ \bibnamefont {Brown}}, \bibinfo {author} {\bibfnamefont {P.~S.}\ \bibnamefont {Crozier}}, \bibinfo {author} {\bibfnamefont {P.~J.}\ \bibnamefont {In't~Veld}}, \bibinfo {author} {\bibfnamefont {A.}~\bibnamefont {Kohlmeyer}}, \bibinfo {author} {\bibfnamefont {S.~G.}\ \bibnamefont {Moore}}, \bibinfo {author} {\bibfnamefont {T.~D.}\ \bibnamefont {Nguyen}}, \emph {et~al.},\ }\href@noop {} {\bibfield  {journal} {\bibinfo  {journal} {Computer Physics Communications}\ }\textbf {\bibinfo {volume} {271}},\ \bibinfo {pages} {108171} (\bibinfo {year} {2022})}\BibitemShut {NoStop}%
	\bibitem [{\citenamefont {Agarwal}\ and\ \citenamefont {Capers}(1972)}]{agarwal1972measurement}%
	  \BibitemOpen
	  \bibfield  {author} {\bibinfo {author} {\bibfnamefont {M.}~\bibnamefont {Agarwal}}\ and\ \bibinfo {author} {\bibfnamefont {M.}~\bibnamefont {Capers}},\ }\href@noop {} {\bibfield  {journal} {\bibinfo  {journal} {Applied Crystallography}\ }\textbf {\bibinfo {volume} {5}},\ \bibinfo {pages} {63} (\bibinfo {year} {1972})}\BibitemShut {NoStop}%
	\bibitem [{\citenamefont {Hamann}(2013)}]{hamann2013optimized}%
	  \BibitemOpen
	  \bibfield  {author} {\bibinfo {author} {\bibfnamefont {D.~R.}\ \bibnamefont {Hamann}},\ }\href@noop {} {\bibfield  {journal} {\bibinfo  {journal} {Physical Review B}\ }\textbf {\bibinfo {volume} {88}},\ \bibinfo {pages} {085117} (\bibinfo {year} {2013})}\BibitemShut {NoStop}%
	\bibitem [{\citenamefont {Fern{\'a}ndez-Seivane}\ \emph {et~al.}(2006)\citenamefont {Fern{\'a}ndez-Seivane}, \citenamefont {Oliveira}, \citenamefont {Sanvito},\ and\ \citenamefont {Ferrer}}]{fernandez2006site}%
	  \BibitemOpen
	  \bibfield  {author} {\bibinfo {author} {\bibfnamefont {L.}~\bibnamefont {Fern{\'a}ndez-Seivane}}, \bibinfo {author} {\bibfnamefont {M.~A.}\ \bibnamefont {Oliveira}}, \bibinfo {author} {\bibfnamefont {S.}~\bibnamefont {Sanvito}},\ and\ \bibinfo {author} {\bibfnamefont {J.}~\bibnamefont {Ferrer}},\ }\href@noop {} {\bibfield  {journal} {\bibinfo  {journal} {Journal of Physics: Condensed Matter}\ }\textbf {\bibinfo {volume} {18}},\ \bibinfo {pages} {7999} (\bibinfo {year} {2006})}\BibitemShut {NoStop}%
	\end{thebibliography}

%

\end{document}